\documentclass[journal]{IEEEtran}

\usepackage{cite}
\usepackage{amsmath,amssymb,amsfonts,amsthm}
\usepackage{algorithmic}
\usepackage{graphicx,xcolor}
\usepackage{textcomp}
\usepackage{cases}
\usepackage[utf8]{inputenc}
\usepackage[english]{babel}
\def\BibTeX{{\rm B\kern-.05em{\sc i\kern-.025em b}\kern-.08em
    T\kern-.1667em\lower.7ex\hbox{E}\kern-.125emX}}

\usepackage{multirow}
\usepackage{pifont}
\newcommand{\cmark}{\ding{52}}  

\usepackage{epstopdf}

\usepackage{enumitem}

\usepackage{hyperref}
\hypersetup{colorlinks=true,linkcolor=black,urlcolor=blue,}

\usepackage{dirtytalk}

\usepackage[caption=false]{subfig} 
\makeatletter
\renewcommand{\fnum@figure}{Fig. \thefigure}
\makeatother

\usepackage[acronym,toc,shortcuts]{glossaries}
\newacronym{Ph.D.}{Ph.D.}{Doctor of Philosophy}
\newacronym{SA}{SA}{Simulated Annealing}
\newacronym{VLC}{VLC}{visible light communications}
\newacronym{RF}{RF}{radio frequency}
\newacronym{V2V}{V2V}{vehicle-to-vehicle}
\newacronym{V2X}{V2X}{vehicle-to-everything}
\newacronym{V2I}{V2I}{vehicle-to-infrastructure}
\newacronym{B5G}{B5G}{beyond-fifth generation}
\newacronym{LED}{LED}{light emitting diode}
\newacronym{OMA}{OMA}{orthogonal multiple access}
\newacronym{FDMA}{FDMA}{frequency-division multiple-access}
\newacronym{TDMA}{TDMA}{time-division multiple-access}
\newacronym{CDMA}{CDMA}{code-division multiple-access}
\newacronym{OFDMA}{OFDMA}{orthogonal frequency-division multiple-access}
\newacronym{OFDM}{OFDM}{orthogonal frequency-division multiplexing}
\newacronym{WDMA}{WDMA}{wavelength-division multiple-access}
\newacronym{NOMA}{NOMA}{non-orthogonal multiple access}
\newacronym{PD-NOMA}{PD-NOMA}{power-domain NOMA}
\newacronym{CD-NOMA}{CD-NOMA}{code-domain NOMA}
\newacronym{SC}{SC}{superposition coding}
\newacronym{SIC}{SIC}{successive interference cancellation}
\newacronym{BS}{BS}{base station}
\newacronym{QoS}{QoS}{quality-of-service}
\newacronym{NP}{NP}{non-deterministic polynomial-time}
\newacronym{DCO-OFDM}{DCO-OFDM}{direct-current biased optical-OFDM}
\newacronym{DCO-OFDMA}{DCO-OFDMA}{direct-current biased optical-OFDMA}
\newacronym{DC}{DC}{direct current}
\newacronym{ITU}{ITU}{international telecommunication union}
\newacronym{FoV}{FoV}{field-of-view}
\newacronym{CSI}{CSI}{channel state information}
\newacronym{LACO-OFDM}{LACO-OFDM}{layered asymmetrically clipped optical OFDM}
\newacronym{ACO-OFDM}{ACO-OFDM}{asymmetrically clipped optical OFDM}
\newacronym{FR}{FR}{frequency reuse}
\newacronym{EAs}{EAs}{evolutionary algorithms}
\newacronym{C-LiAN}{C-LiAN}{centralized light access network}
\newacronym{AP}{AP}{access point}
\newacronym{PD}{PD}{photo-diode}
\newacronym{SINR}{SINR}{signal-to-noise-interference ratio}
\newacronym{LoS}{LoS}{line-of-sight}
\newacronym{AWGN}{AWGN}{additive white Gaussian noise}
\newacronym{SNR}{SNR}{signal-to-noise ratio}
\newacronym{NLUPA}{NLUPA}{next-largest-difference user-pairing algorithm}
\newacronym{D-NLUPA}{D-NLUPA}{divide-and-next-largest-difference user-pairing algorithm}
\newacronym{NAICS}{NAICS}{network-assisted interference cancellation and suppression}
\newacronym{LTE}{LTE}{long term evolution}
\newacronym{3GPP}{3GPP}{3rd Generation Partnership Project}
\newacronym{CR}{CR}{cognitive radio}
\newacronym{2D}{2D}{two-dimension}
\newacronym{GP}{GP}{gradient projection}
\newacronym{umMTC}{umMTC}{ultra-massive machine-type communication}
\newacronym{mMTC}{mMTC}{massive machine-type communication}
\newacronym{IoE}{IoE}{internet-of-everything}
\newacronym{IoUT}{IoUT}{internet-of-underwater-things}
\newacronym{ZF}{ZF}{zero-forcing}
\newacronym{NLIP}{NLIP}{non-linear integer programming}
\newacronym{DP}{DP}{dynamic programming}
\newacronym{MIMO}{MIMO}{multiple-input multiple-output}
\newacronym{TS}{TS}{Tabu-search}
\newacronym{THz}{THz}{Terahertz}
\newacronym{MISO}{MISO}{multiple-input single-output}
\newacronym{SIMO}{SIMO}{single-input multiple-output}
\newacronym{EE}{EE}{energy efficiency}
\newacronym{SEE}{SEE}{sum energy efficiency}
\newacronym{VR}{VR}{virtual reality}
\newacronym{XR}{XR}{extended reality}
\newacronym{5G}{5G}{fifth generation}
\newacronym{6G}{6G}{sixth generation}
\newacronym{NR}{NR}{new radio}
\newacronym{mmWave}{mmWave}{millimeter-wave}
\newacronym{FD}{FD}{full-duplex}
\newacronym{CNOMA}{CNOMA}{cooperative NOMA}
\newacronym{ABF}{ABF}{analog beamforming}
\newacronym{BF}{BF}{beamforming}
\newacronym{DF}{DF}{decode-and-forward}
\newacronym{AF}{AF}{amplify-and-forward}
\newacronym{CF}{CF}{compress-and-forward}
\newacronym{SPS}{SPS}{single-phase shifter}
\newacronym{PS}{PS}{phase shifter}
\newacronym{PA}{PA}{power amplifier}
\newacronym{NLoS}{NLoS}{non-line-of-sight}
\newacronym{ULA}{ULA}{uniform linear array}
\newacronym{SI}{SI}{self-interference}
\newacronym{MA}{MA}{multiple access}
\newacronym{1G}{1G}{first generation}
\newacronym{2G}{2G}{second generation}
\newacronym{3G}{3G}{third generation}
\newacronym{4G}{4G}{fourth generation}
\newacronym{M2M}{M2M}{machine-to-machine}
\newacronym{IoT}{IoT}{internet-of-things}
\newacronym{IMT}{IMT}{International Mobile Telecommunications}
\newacronym{SE}{SE}{spectral efficiency}
\newacronym{MMF}{MMF}{Maximin Fairness}
\newacronym{SR}{SR}{sum rate}
\newacronym{WSR}{WSR}{weighted sum rate}
\newacronym{D-NOMA}{D-NOMA}{dynamic-NOMA}
\newacronym{GSM}{GSM}{global system for mobile telecommunications}
\newacronym{IS-95}{IS-95}{Interim Standard 95}
\newacronym{SMS}{SMS}{short-message service}
\newacronym{CoMP}{CoMP}{coordinated multi-point}
\newacronym{MU-MIMO}{MU-MIMO}{multi-user MIMO}
\newacronym{MAC}{MAC}{multiple-access channel}
\newacronym{BC}{BC}{vector-broadcast channel}
\newacronym{CSIT}{CSIT}{channel state information at the transmitter}
\newacronym{DPC}{DPC}{dirty paper coding}
\newacronym{ZF-DPC}{ZF-DPC}{zero-forcing DPC}
\newacronym{BD}{BD}{block-diagonalization}
\newacronym{ICI}{ICI}{inter-cell interference}
\newacronym{MMSE}{MMSE}{minimum mean square error}
\newacronym{LTE-A}{LTE-A}{long term evolution-advanced}
\newacronym{MUST}{MUST}{multi-user superposition transmission}
\newacronym{SISO}{SISO}{single-input single-output}
\newacronym{LDS-CDMA}{LDS-CDMA}{low-density spreading CDMA}
\newacronym{LDS-OFDM}{LDS-OFDM}{low-density spreading OFDM}
\newacronym{SCMA}{SCMA}{sparse code multiple access}
\newacronym{MUSA}{MUSA}{multi-user sharing access}
\newacronym{SAMA}{SAMA}{successive interference cancellation amenable multiple access}
\newacronym{PDMA}{PDMA}{pattern division multiple access} 
\newacronym{BOMA}{BOMA}{building block sparse-constellation based orthogonal multiple access}  
\newacronym{LPMA}{LPMA}{lattice partition multiple access} \newacronym{OOK}{OOK}{on-off keying} 
\newacronym{M-PAM}{M-PAM}{M-ary pulse-amplitude modulation} 
\newacronym{M-PPM}{M-PPM}{M-ary pulse-position modulation} \newacronym{MSM}{MSM}{multiple-subcarrier modulation}
\newacronym{IM/DD}{IM/DD}{intensity modulation and direct detection}
\newacronym{RC}{RC}{repetition code}
\newacronym{SM}{SM}{spatial multiplexing}
\newacronym{SMOD}{SMOD}{spatial modulation}
\newacronym{BER}{BER}{bit error rate}
\newacronym{SER}{SER}{symbol error rate}
\newacronym{MFTP}{MFTP}{maximum flickering time period}
\newacronym{MU-MISO}{MU-MISO}{multi-user multi-input single-output}
\newacronym{MSE}{MSE}{minimum square error}
\newacronym{SPCA}{SPCA}{sequential parametric convex approximation}
\newacronym{WSMSE}{WSMSE}{weighted sum minimum square error}
\newacronym{OCDMA}{OCDMA}{optical code division multiple access}
\newacronym{SDMA}{SDMA}{space-division multiple access}
\newacronym{DMT}{DMT}{discrete multi-tone}
\newacronym{VLNs}{VLNs}{visible light networks}
\newacronym{VHO}{VHO}{Vertical handover}
\newacronym{RSS}{RSS}{received signal strength}
\newacronym{RSI}{RSI}{received signal intensity}
\newacronym{IA}{IA}{interference alignment}
\newacronym{BIA}{BIA}{blind interference alignment}
\newacronym{BBU}{BBU}{base-band unit}
\newacronym{SU}{SU}{secondary user}
\newacronym{PU}{PU}{primary user}
\newacronym{mMIMO}{mMIMO}{massive-MIMO}
\newacronym{UAV}{UAV}{unmanned aerial vehicle}
\newacronym{PHY}{PHY}{physical}
\newacronym{CF-mMIMO}{CF-mMIMO}{cell-free mMIMO}
\newacronym{LIS}{LIS}{large intelligent surfaces}
\newacronym{3-D MIMO}{3-D MIMO}{3-Dimensional MIMO}
\newacronym{RIS}{RIS}{reflecting intelligent surface}
\newacronym{BackCom}{BackCom}{backscatter communications}
\newacronym{UL}{UL}{uplink}
\newacronym{UE}{UE}{user equipment}
\newacronym{D2D}{D2D}{device-to-device}
\newacronym{FCC}{FCC}{Federal Communications Commission}
\newacronym{HAP}{HAP}{high altitude platform}
\newacronym{LAP}{LAP}{low altitude platform}
\newacronym{MEC}{MEC}{mobile edge computing}
\newacronym{NATO}{NATO}{North Atlantic Treaty Organization}
\newacronym{ML}{ML}{machine learning}
\newacronym{QML}{QML}{quantum machine learning}
\newacronym{DL}{DL}{deep learning}
\newacronym{DRL}{DRL}{deep Reinforcement learning}
\newacronym{RL}{RL}{Reinforcement learning}
\newacronym{MMA}{MMA}{minorization maximization algorithm}
\newacronym{MM}{MM}{majorization-minimization}
\newacronym{KKT}{KKT}{Karush–Kuhn–Tucker}
\newacronym{FDD}{FDD}{frequency division duplex}
\newacronym{SCA}{SCA}{sine-cosine algorithm}
\newacronym{AoD}{AoD}{angle-of-departure}
\newacronym{SDP}{SDP}{semi-definite programming}
\newacronym{SDR}{SDR}{semi-definite relaxation}
\newacronym{JT}{JT}{joint transmission}
\newacronym{CB}{CB}{coordinated beamforming}
\newacronym{RAMA}{RAMA}{relay-aided multiple access}
\newacronym{MRC}{MRC}{maximum ratio combining}
\newacronym{SWIPT}{SWIPT}{simultaneous wireless information and power transfer}
\newacronym{HetNets}{HetNets}{heterogeneous networks}
\newacronym{D.C.}{D.C.}{difference of convex}
\newacronym{GRPA}{GRPA}{gain ratio power allocation}
\newacronym{FPA}{FPA}{fixed power allocation}
\newacronym{CS}{CS}{Cuckoo Search}
\newacronym{HHO}{HHO}{Harris Hawks Optimizer}
\newacronym{PLC}{PLC}{power line communications}
\newacronym{HTT}{HTT}{harvest-then-transmit}
\newacronym{H-CRAN}{H-CRAN}{heterogeneous cloud radio access network}
\newacronym{RRHs}{RRHs}{remote radio heads}
\newacronym{IIoT}{IIoT}{Industrial IoT}
\newacronym{PAPR}{PAPR}{peak-to-average-power-ratio}
\newacronym{ANC}{ANC}{analog network coding}
\newacronym{FFR}{FFR}{fractional frequency reuse}
\newacronym{RGB}{RGB}{red-green-blue}
\newacronym{MAR}{MAR}{mobile augmented reality}
\newacronym{HD}{HD}{half-duplex}
\newacronym{CPU}{CPU}{central process unit}
\newacronym{RWP}{RWP}{Random Way-Point}
\newacronym{LC}{LC}{liquid crystal}
\newacronym{ADR}{ADR}{angle diversity receiver}
\newacronym{RSMA}{RSMA}{rate splitting multiple access}
\newacronym{omni-DRIS}{omni-DRIS}{omni-digital-RIS}
\newacronym{STAR-RIS}{STAR-RIS}{simultaneous transmission and reflection reconfigurable intelligent surface}
\newacronym{OSTAR-RIS}{OSTAR-RIS}{optical simultaneous transmission and reflection reconfigurable intelligent surface}
\glsdisablehyper

\usepackage[linesnumbered,ruled,vlined]{algorithm2e}

\begin{document}

\title{Optical STAR-RIS-Aided VLC Systems: RSMA Versus NOMA}

\author{Omar~Maraqa,~Sylvester~Aboagye,~\IEEEmembership{Member,~IEEE}, and Telex~M.~N.~Ngatched, \IEEEmembership{Senior~Member,~IEEE}%

\thanks{Omar~Maraqa and Telex~M.~N.~Ngatched are with the Department of Electrical and Computer Engineering, McMaster University, Hamilton, Canada (e-mail: dr.omar.maraqa@gmail.com; ngatchet@mcmaster.ca). Sylvester~Aboagye is with the Department of Electrical Engineering and Computer Science, York University, Toronto, Canada (e-mail: aboagye@yorku.ca).}%
\thanks{$\copyright$ 2023 IEEE. Personal use of this material is permitted. Permission from IEEE must be obtained for all other uses, in any current or future media, including reprinting/republishing this material for advertising or promotional purposes, creating new collective works, for resale or redistribution to servers or lists, or reuse of any copyrighted component of this work in other works.}%
\thanks{Digital Object Identifier:\href{https://ieeexplore.ieee.org/abstract/document/10375270}{10.1109/OJCOMS.2023.3347534}}
}

\markboth{Accepted for publication in IEEE Open Journal of the Communications Society, Dec. 2023.}%
{Maraqa \MakeLowercase{\textit{et al.}}: Optical STAR-RIS-Aided VLC systems: RSMA versus NOMA}

\maketitle

\begin{abstract}
A critical concern within the realm of visible light communications (VLC) pertains to enhancing system data rate, particularly in scenarios where the direct line-of-sight (LoS) connection is obstructed by obstacles. The deployment of meta-surface-based simultaneous transmission and reflection reconfigurable intelligent surface (STAR-RIS) has emerged to combat challenging LoS blockage scenarios and to provide $360^{\circ}$ coverage in radio-frequency wireless systems. Recently, the concept of optical simultaneous transmission and reflection reconfigurable intelligent surface (OSTAR-RIS) has been promoted for VLC systems. This work is dedicated to studying the performance of OSTAR-RIS in detail and unveiling the VLC system performance gain under such technology. Specifically, we propose a novel multi-user indoor VLC system that is assisted by OSTAR-RIS. To improve the sum rate performance of the proposed system, both power-domain non-orthogonal multiple access (NOMA) and rate splitting multiple access (RSMA) are investigated in this work. To realize this, a sum rate maximization problem that jointly optimizes the roll and yaw angles of the reflector elements as well as the refractive index of the refractor elements in OSTAR-RIS is formulated, solved, and evaluated. The maximization problem takes into account practical considerations, such as the presence of non-users (i.e., blockers) and the orientation of the recipient's device. The sine-cosine meta-heuristic algorithm is employed to get the optimal solution of the formulated non-convex optimization problem. Moreover, the study delves into the sum energy efficiency optimization of the proposed system. Simulation results indicate that the proposed OSTAR-RIS RSMA-aided VLC system outperforms the OSTAR-RIS NOMA-based VLC system in terms of both the sum rate and the sum energy efficiency.        
\end{abstract}


\maketitle

\section{Introduction}

\IEEEPARstart{T}{he} rapid growth in the number of interconnected devices and the continuous advancement of wireless applications are prompting the exploration of alternative wireless communication solutions beyond \gls{RF} communications. \Gls{VLC} is a technology that uses visible light to communicate, and has many advantages such as being cost-efficient, bandwidth-abundant, and highly-secured~\cite{8528460}. Consequently, \gls{VLC} is regarded as a promising complementary technology to \gls{RF} communications within upcoming wireless networks~\cite{8528460}.

\begin{table*}[!t]
\centering
\vspace{-2.0em}
\caption{State-of-the-art RIS-aided optimized solutions in VLC systems. (``ASNR'':``Average signal-to-noise ratio'', ``BER'':``bit error rate'', ``MSE'':``mean square error'', ``E'':``Energy'', and ``EE'':``Energy efficiency'')}
\label{tab: Related work}
\resizebox{\textwidth}{!}{%
\begin{tabular}{|c|cc|cc|ccccccc|ccc|}
\hline
\multirow{2}{*}{\textbf{{[}\#{]}}} & \multicolumn{4}{c|}{\textbf{RIS type}} & \multicolumn{7}{c|}{\textbf{Objective function}} & \multicolumn{3}{c|}{\textbf{Access technique}} \\ \cline{2-15} 

 & \multicolumn{1}{c|}{\textbf{\begin{tabular}[c]{@{}c@{}}Meta-\\surface-\\ based\\ RIS\end{tabular}}} & \textbf{\begin{tabular}[c]{@{}c@{}}Mirror \\array-\\ based\\ RIS\end{tabular}} & \multicolumn{1}{c|}{\textbf{\begin{tabular}[c]{@{}c@{}} \\ STAR-RIS \\ (Meta-surface-\\ based RIS)\end{tabular}}} & \textbf{\begin{tabular}[c]{@{}c@{}} \\ OSTAR-RIS \\ (Liquid crystal \\ and \\ mirror array-\\ based RIS)\end{tabular}} & \multicolumn{1}{c|}{\textbf{ASNR}} & \multicolumn{1}{c|}{\textbf{MSE}} & \multicolumn{1}{c|}{\textbf{BER}} & \multicolumn{1}{c|}{\textbf{\begin{tabular}[c]{@{}c@{}}Secrecy\\ Rate\end{tabular}}} & \multicolumn{1}{c|}{\textbf{E}} & \multicolumn{1}{c|}{\textbf{Rate}} & \textbf{\begin{tabular}[c]{@{}c@{}} EE \end{tabular}} & \multicolumn{1}{c|}{\textbf{OMA}} & \multicolumn{1}{c|}{\textbf{NOMA}} & \textbf{RSMA} \\ \hline

\cite{9543660} & \multicolumn{1}{c|}{} & \cmark & \multicolumn{1}{c|}{} &  & \multicolumn{1}{c|}{} & \multicolumn{1}{c|}{} & \multicolumn{1}{c|}{} & \multicolumn{1}{c|}{} & \multicolumn{1}{c|}{} & \multicolumn{1}{c|}{\cmark} &  & \multicolumn{1}{c|}{\cmark} & \multicolumn{1}{c|}{} &  \\ \hline

\cite{9526581} & \multicolumn{1}{c|}{} & \cmark & \multicolumn{1}{c|}{} &  & \multicolumn{1}{c|}{} & \multicolumn{1}{c|}{} & \multicolumn{1}{c|}{} & \multicolumn{1}{c|}{} & \multicolumn{1}{c|}{} & \multicolumn{1}{c|}{\cmark} &  & \multicolumn{1}{c|}{\cmark} & \multicolumn{1}{c|}{} &  \\ \hline

\cite{9500409} & \multicolumn{1}{c|}{} & \cmark & \multicolumn{1}{c|}{} &  & \multicolumn{1}{c|}{} & \multicolumn{1}{c|}{} & \multicolumn{1}{c|}{} & \multicolumn{1}{c|}{\cmark} & \multicolumn{1}{c|}{} & \multicolumn{1}{c|}{} &  & \multicolumn{1}{c|}{\cmark} & \multicolumn{1}{c|}{} &  \\ \hline

\cite{abumarshoud2022intelligent} & \multicolumn{1}{c|}{} & \cmark & \multicolumn{1}{c|}{} &  & \multicolumn{1}{c|}{} & \multicolumn{1}{c|}{} & \multicolumn{1}{c|}{} & \multicolumn{1}{c|}{\cmark} & \multicolumn{1}{c|}{} & \multicolumn{1}{c|}{} &  & \multicolumn{1}{c|}{} & \multicolumn{1}{c|}{\cmark} &  \\ \hline

\cite{9714890} & \multicolumn{1}{c|}{} & \cmark & \multicolumn{1}{c|}{} &  & \multicolumn{1}{c|}{} & \multicolumn{1}{c|}{} & \multicolumn{1}{c|}{} & \multicolumn{1}{c|}{} & \multicolumn{1}{c|}{} & \multicolumn{1}{c|}{\cmark} &  & \multicolumn{1}{c|}{\cmark} & \multicolumn{1}{c|}{} &  \\ \hline

\cite{10024150} & \multicolumn{1}{c|}{} &  \cmark & \multicolumn{1}{c|}{} &  & \multicolumn{1}{c|}{} & \multicolumn{1}{c|}{\cmark} & \multicolumn{1}{c|}{} & \multicolumn{1}{c|}{} & \multicolumn{1}{c|}{} & \multicolumn{1}{c|}{} &  & \multicolumn{1}{c|}{\cmark} & \multicolumn{1}{c|}{} &  \\ \hline

\cite{wu2022configuring} & \multicolumn{1}{c|}{} & \cmark & \multicolumn{1}{c|}{} &  & \multicolumn{1}{c|}{} & \multicolumn{1}{c|}{} & \multicolumn{1}{c|}{} & \multicolumn{1}{c|}{} & \multicolumn{1}{c|}{} & \multicolumn{1}{c|}{\cmark} &  & \multicolumn{1}{c|}{\cmark} & \multicolumn{1}{c|}{} &  \\ \hline

\cite{shi2022performance} & \multicolumn{1}{c|}{} & \cmark & \multicolumn{1}{c|}{} &  & \multicolumn{1}{c|}{} & \multicolumn{1}{c|}{} & \multicolumn{1}{c|}{\cmark} & \multicolumn{1}{c|}{} & \multicolumn{1}{c|}{} & \multicolumn{1}{c|}{} &  & \multicolumn{1}{c|}{\cmark} & \multicolumn{1}{c|}{} &  \\ \hline

\cite{yang2022average} & \multicolumn{1}{c|}{} & \cmark & \multicolumn{1}{c|}{} &  & \multicolumn{1}{c|}{\cmark} & \multicolumn{1}{c|}{} & \multicolumn{1}{c|}{} & \multicolumn{1}{c|}{} & \multicolumn{1}{c|}{} & \multicolumn{1}{c|}{} &  & \multicolumn{1}{c|}{\cmark} & \multicolumn{1}{c|}{} &  \\ \hline

\cite{9784887} & \multicolumn{1}{c|}{} & \cmark & \multicolumn{1}{c|}{} &  & \multicolumn{1}{c|}{} & \multicolumn{1}{c|}{} & \multicolumn{1}{c|}{} & \multicolumn{1}{c|}{\cmark} & \multicolumn{1}{c|}{} & \multicolumn{1}{c|}{} &  & \multicolumn{1}{c|}{\cmark} & \multicolumn{1}{c|}{} &  \\ \hline

\cite{9756553} & \multicolumn{1}{c|}{} & \cmark & \multicolumn{1}{c|}{} &  & \multicolumn{1}{c|}{} & \multicolumn{1}{c|}{} & \multicolumn{1}{c|}{} & \multicolumn{1}{c|}{\cmark} & \multicolumn{1}{c|}{} & \multicolumn{1}{c|}{} &  & \multicolumn{1}{c|}{\cmark} & \multicolumn{1}{c|}{} &  \\ \hline

\cite{zhan2022optimal} & \multicolumn{1}{c|}{} & \cmark & \multicolumn{1}{c|}{} &  & \multicolumn{1}{c|}{} & \multicolumn{1}{c|}{} & \multicolumn{1}{c|}{} & \multicolumn{1}{c|}{} & \multicolumn{1}{c|}{} & \multicolumn{1}{c|}{} & \cmark & \multicolumn{1}{c|}{\cmark} & \multicolumn{1}{c|}{} &  \\ \hline

\cite{wu2022capacity} & \multicolumn{1}{c|}{} & \cmark & \multicolumn{1}{c|}{} &  & \multicolumn{1}{c|}{} & \multicolumn{1}{c|}{} & \multicolumn{1}{c|}{} & \multicolumn{1}{c|}{} & \multicolumn{1}{c|}{} & \multicolumn{1}{c|}{\cmark} &  & \multicolumn{1}{c|}{\cmark} & \multicolumn{1}{c|}{} &  \\ \hline

\cite{wu2022position} & \multicolumn{1}{c|}{} & \cmark & \multicolumn{1}{c|}{} &  & \multicolumn{1}{c|}{} & \multicolumn{1}{c|}{} & \multicolumn{1}{c|}{} & \multicolumn{1}{c|}{} & \multicolumn{1}{c|}{} & \multicolumn{1}{c|}{\cmark} &  & \multicolumn{1}{c|}{\cmark} & \multicolumn{1}{c|}{} &  \\ \hline

\cite{10183987} & \multicolumn{1}{c|}{} & \cmark & \multicolumn{1}{c|}{} &  & \multicolumn{1}{c|}{} & \multicolumn{1}{c|}{} & \multicolumn{1}{c|}{} & \multicolumn{1}{c|}{} & \multicolumn{1}{c|}{} & \multicolumn{1}{c|}{\cmark} & \cmark & \multicolumn{1}{c|}{\cmark} & \multicolumn{1}{c|}{\cmark} &  \\ \hline

\cite{10190313} & \multicolumn{1}{c|}{\cmark} &  & \multicolumn{1}{c|}{} &  & \multicolumn{1}{c|}{} & \multicolumn{1}{c|}{} & \multicolumn{1}{c|}{} & \multicolumn{1}{c|}{} & \multicolumn{1}{c|}{} & \multicolumn{1}{c|}{\cmark} &  & \multicolumn{1}{c|}{\cmark} & \multicolumn{1}{c|}{} &  \\ \hline

\cite{9348585} & \multicolumn{1}{c|}{\cmark} &  & \multicolumn{1}{c|}{} &  & \multicolumn{1}{c|}{} & \multicolumn{1}{c|}{} & \multicolumn{1}{c|}{} & \multicolumn{1}{c|}{} & \multicolumn{1}{c|}{} & \multicolumn{1}{c|}{} & \cmark & \multicolumn{1}{c|}{\cmark} & \multicolumn{1}{c|}{} &  \\ \hline

\cite{10008547} & \multicolumn{1}{c|}{\cmark} &  & \multicolumn{1}{c|}{} &  & \multicolumn{1}{c|}{} & \multicolumn{1}{c|}{} & \multicolumn{1}{c|}{} & \multicolumn{1}{c|}{} & \multicolumn{1}{c|}{} & \multicolumn{1}{c|}{\cmark} &  & \multicolumn{1}{c|}{\cmark} & \multicolumn{1}{c|}{\cmark} &  \\ \hline

\cite{9838853} & \multicolumn{1}{c|}{\cmark} &  & \multicolumn{1}{c|}{} &  & \multicolumn{1}{c|}{} & \multicolumn{1}{c|}{} & \multicolumn{1}{c|}{\cmark} & \multicolumn{1}{c|}{} & \multicolumn{1}{c|}{} & \multicolumn{1}{c|}{} &  & \multicolumn{1}{c|}{} & \multicolumn{1}{c|}{\cmark} &  \\ \hline

\cite{10047999} & \multicolumn{1}{c|}{\cmark} &  & \multicolumn{1}{c|}{} &  & \multicolumn{1}{c|}{} & \multicolumn{1}{c|}{} & \multicolumn{1}{c|}{} & \multicolumn{1}{c|}{} & \multicolumn{1}{c|}{} & \multicolumn{1}{c|}{\cmark} &  & \multicolumn{1}{c|}{} & \multicolumn{1}{c|}{\cmark} &  \\ \hline

\cite{10236455} & \multicolumn{1}{c|}{\cmark} &  & \multicolumn{1}{c|}{} &  & \multicolumn{1}{c|}{} & \multicolumn{1}{c|}{} & \multicolumn{1}{c|}{} & \multicolumn{1}{c|}{} & \multicolumn{1}{c|}{\cmark} & \multicolumn{1}{c|}{} &  & \multicolumn{1}{c|}{\cmark} & \multicolumn{1}{c|}{} &  \\ \hline

\cite{salehiyan2022performance} & \multicolumn{1}{c|}{} &  & \multicolumn{1}{c|}{\cmark} &  & \multicolumn{1}{c|}{} & \multicolumn{1}{c|}{} & \multicolumn{1}{c|}{} & \multicolumn{1}{c|}{} & \multicolumn{1}{c|}{} & \multicolumn{1}{c|}{\cmark} & \cmark & \multicolumn{1}{c|}{} & \multicolumn{1}{c|}{\cmark} &  \\ \hline

\begin{tabular}[c]{@{}c@{}}Proposed \\ solution\end{tabular}& \multicolumn{1}{c|}{} &  & \multicolumn{1}{c|}{} & \cmark & \multicolumn{1}{c|}{} & \multicolumn{1}{c|}{} & \multicolumn{1}{c|}{} & \multicolumn{1}{c|}{} & \multicolumn{1}{c|}{} & \multicolumn{1}{c|}{\cmark} & \cmark & \multicolumn{1}{c|}{} & \multicolumn{1}{c|}{\cmark} & \cmark \\ \hline
\end{tabular}%
}
\end{table*}

Reliable data transmission in \gls{VLC} systems requires a clear \gls{LoS} path between the transmitter and the receiver~\cite{9543660}. Nevertheless, this direct path can be blocked due to factors such as the presence of other users (i.e., blockers), obstacles within indoor environments, or the orientation of the recipient's device. Many previous investigations into \gls{VLC} systems neglected to account for the impact of random device orientations, even though this is a crucial consideration for practical \gls{VLC} system design and evaluation. 

In another development, in the last couple of years, researchers have proposed the deployment of \glspl{RIS} on walls to enhance the quality of the communication links in \gls{VLC} systems and to combat the \gls{LoS} blockage. In instances where the \gls{LoS} path is obstructed, an \gls{RIS} can steer the light propagation in the wireless channel to overcome the blockage. Therefore, the integration of \gls{RIS} technology can reduce the reliance of \gls{VLC} systems on \gls{LoS} paths. Two main streams have evolved concerning the adoption of \gls{RIS} in \gls{VLC} systems, namely, mirror array-based \gls{RIS}~\cite{9526581,9543660,9500409,abumarshoud2022intelligent,9714890,10024150,wu2022configuring,shi2022performance,yang2022average,9784887,9756553, zhan2022optimal, wu2022capacity, wu2022position,10183987} and meta-surface-based \gls{RIS}~\cite{10190313,9348585,10008547,9838853,10047999,10236455}. The authors in~\cite{9276478} observed that the mirror array-based \gls{RIS} demonstrates superior performance gains compared to the meta-surface-based \gls{RIS}.

To provide $360^{\circ}$ coverage and to combat challenging \gls{LoS} blockage scenarios, meta-surface-based \gls{STAR-RIS} was investigated for \gls{VLC} systems in~\cite{salehiyan2022performance}. As a result of the more recent developments in the \gls{STAR-RIS} materials, the concept and the fabrication method of \gls{OSTAR-RIS} are presented in~\cite{ndjiongue2022double}. The proposed \gls{OSTAR-RIS} consists of \gls{LC}-based \gls{RIS} that can be designed to act as reflectors or/and refractors.

On the other hand, in a separate development, a new interference management approach referred to as \gls{RSMA}~\cite{9831440,10038476} has been gaining attention for its ability to enhance the spectral efficiency within a wide range of interference scenarios. \gls{RSMA} has been demonstrated to embrace and outperform existing multiple access schemes, namely, \gls{OMA}, space division multiple access based on linear precoding, physical layer multicasting, and \gls{NOMA} based on linear superposition coding with \gls{SIC}~\cite{10038476}. 

\begin{figure*}[!t]
\centering
\vspace{-2em}
\includegraphics[width=0.8\textwidth]{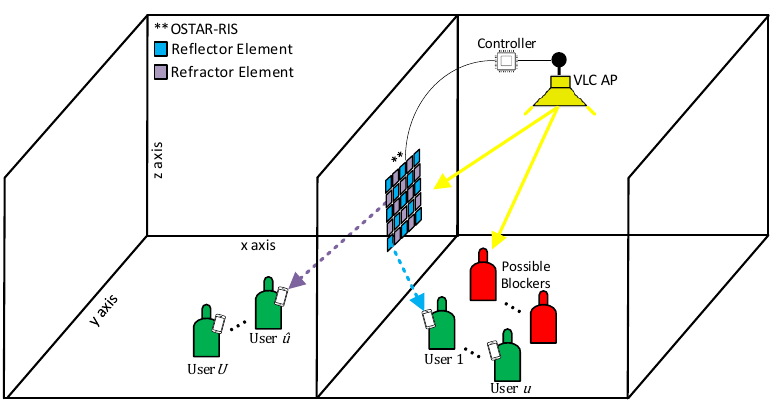}
\caption{An illustration of the proposed \gls{OSTAR-RIS}-aided \gls{VLC} system.}
\label{fig: LC-STAR-RIS-Aided system model}
\vspace{-1.5em}
\end{figure*}

In this work, a novel multi-user indoor \gls{VLC} system that is assisted by \gls{OSTAR-RIS} is presented, optimized, and evaluated. Unlike the design in~\cite{ndjiongue2022double}, our \gls{OSTAR-RIS} is made of both \gls{LC}-based \gls{RIS} and mirror-array based \gls{RIS}. The choice of utilizing mirror array-based \gls{RIS} elements as reflectors instead of meta-surface-based \gls{RIS} elements arises from the superior performance gain the former can provide compared to the latter, as observed in~\cite{9276478}. The choice of utilizing the \gls{LC}-based \gls{RIS} elements as refractors arises from the fact that the \gls{LC}-based \gls{RIS} element has electrically controllable birefringence, which can smoothly steer and amplify the light's signal through the \gls{LC} substance~\cite{ndjiongue2022double}. These joint signal steering and amplification capabilities, introduce new possibilities in the design of \gls{RIS}-aided \gls{VLC} systems compared to meta-surface-based \glspl{RIS} that can only control the direction of the refracted signal. For the reader's convenience, Table~\ref{tab: Related work} summarizes the most related works and compares them to our proposed work according to (i) the adopted \gls{RIS} type, (ii) the optimized objective function, and (iii) the adopted multiple access technique. From this table, it is visible that most of the works have considered \gls{OMA}-based schemes, some works have considered power-domain \gls{NOMA} scheme to boost the achievable rate in \gls{VLC} systems, but none has considered the \gls{RSMA} scheme. To the best of the authors' knowledge, this is the first work that technically evaluates \gls{OSTAR-RIS} under the \gls{RSMA} scheme. Overall, the novelty in this manuscript is two-fold, first, in proposing and investigating a new \gls{STAR-RIS} design, and second, in evaluating the \gls{RSMA} scheme in \gls{RIS}-enabled \gls{VLC} systems.

Our contributions can be listed as follows:
\begin{itemize}[noitemsep,topsep=0pt]
    \item A novel multi-user indoor \gls{VLC} system that is assisted by \gls{OSTAR-RIS} is proposed. The proposed system accounts for both the effect of the orientation of the user's device and the non-user blockers.
    \item A sum rate optimization problem for both the power-domain \gls{NOMA} scheme and the \gls{RSMA} scheme is formulated while jointly designing the yaw and roll angles of the mirror array-based \gls{OSTAR-RIS} elements as well as the refractive index of the \gls{LC}-based \gls{OSTAR-RIS} elements. The sine-cosine meta-heuristic algorithm is utilized to get the global optimal solution for the formulated multi-variate non-convex optimization problem.
    \item A \gls{SEE} optimization problem for both the power-domain \gls{NOMA} scheme and the \gls{RSMA} scheme is also formulated and solved, given the significance of this metric in assessing \gls{VLC} systems.
    \item Detailed simulation results are provided to demonstrate the superiority of the \gls{RSMA} scheme over the power-domain \gls{NOMA} scheme for the proposed system in terms of both the sum rate and the \gls{SEE} metrics while considering different network parameters, such as the \gls{AP} optical transmit power, the wavelength of the transmitted light, the number of elements in the \gls{OSTAR-RIS}, and the number of served users. In addition, the effect of adopting different power allocation strategies on the sum rate performance of the \gls{RSMA} scheme is illustrated. 
\end{itemize}

The organization of the rest of this paper is as follows. In Section~\ref{Sec: System and Channel Models}, the system and channel models of the proposed \gls{OSTAR-RIS}-aided \gls{VLC} system that considers both \gls{LC}-based \gls{OSTAR-RIS} elements and mirror array-based \gls{OSTAR-RIS} elements are presented. The analysis of both adopted multiple access schemes (i.e., power-domain \gls{NOMA} scheme and \gls{RSMA} scheme) are provided in Section~\ref{Sec: Adopted Multiple Access Schemes}. In Section~\ref{Sec: Sum Rate Optimization}, the sum rate maximization problem for both aforementioned multiple access schemes is formulated and optimally solved using the sine-cosine meta-heuristic algorithm. In Section~\ref{Sec: Sum Energy Efficiency Optimization}, a \gls{SEE} optimization of the proposed system is given. Detailed Simulation results are provided in Section~\ref{Sec: Simulation Results}. Finally, the paper's conclusion and future research directions are given in Section~\ref{Sec: Conclusion}.

\section{System and Channel Models}
\label{Sec: System and Channel Models}

In this section, the system and channel models of the proposed \gls{OSTAR-RIS}-aided \gls{VLC} system are presented.

\subsection{Indoor VLC System}
\label{subSec: Indoor VLC System}

An indoor downlink \gls{VLC} system with a single \gls{LED} array as the \gls{AP} and multiple users is illustrated in Fig.~\ref{fig: LC-STAR-RIS-Aided system model}. In this figure, the indoor environment is partitioned into two compartments, with multiple users deployed on each side. Due to room design requirements (e.g., functional design where one compartment is intended as a workspace and the other for relaxation or sleeping, architectural constraints, and individual preferences), one compartment has an \gls{LED} array installed while the other does not have any light fixture. Multiple non-users present in the indoor environment may block the \gls{LoS} signal from the \gls{VLC} \gls{AP} (i.e., non-user blockers). Moreover, we consider self-blockage, which occurs when (i) a user's body blocks the light path to the receiving device or (ii) with random device orientation. To enable wireless data transmission in the compartment without the \gls{LED} array, support non-\gls{LoS} transmissions, and overcome the impact of device orientation on the achievable data rate, an \gls{OSTAR-RIS} is deployed on the wall separating the two compartments. 

The \gls{OSTAR-RIS} is composed of mirror array-based \glspl{RIS} and \gls{LC}-based \glspl{RIS}, that have been interleaved as shown in Fig.~\ref{fig: LC-STAR-RIS-Aided system model}. The mirror array \glspl{RIS} act as reflector elements that reflect and steer incident light to the users. The \gls{LC}-based \glspl{RIS} serve as refractor elements that amplify, refract/transmit, and steer any incident signal to users in the other compartment. The purple dotted and blue dotted lines in the figure denote the refracted light signal and the reflected signal, respectively.

\subsection{VLC Channel}
The channel gain between the \gls{AP} and the users can be expressed as 
\begin{equation} \label{eq: Total Channel Gain}
H= \begin{cases} \iota H_{\textnormal{LoS}} + \sum_{k=1}^\mathcal{K} H_{\textnormal{NLoS}}^{\textnormal{RIS}_k}, \ \textnormal{if the user lies in Room 1} \\ 
\sum_{n=1}^\mathcal{N} H_{\textnormal{NLoS}}^{\textnormal{RIS}_n} \times \psi_{\textnormal{LC}}, \ \ \ \textnormal{if the user lies in Room 2} 
\end{cases}
\end{equation}
\noindent where ``Room 1" and ``Room 2" denote the compartments with and without the \gls{VLC} \gls{AP}, respectively, $\iota\in\{0,1\}$ is an indicator function for the presence or absence of a \gls{LoS} path, $H_{\rm LoS}$ represent the channel gain of the \gls{LoS} path, $H_{\rm NLoS}^{{\rm RIS}_k}$ is the \gls{NLoS} channel gain of the reflected signal from the $k$-th reflector element, $H_{\rm NLoS}^{{\rm RIS}_n}$ represents the \gls{NLoS} channel gain of the refracted signal from the $n$-th refractor element, $\mathcal{K}$ denotes the number of mirror array-based \glspl{RIS}, $\mathcal{N}$ denotes the number of \gls{LC}-based \glspl{RIS}, and $\psi_{\textnormal{LC}}$ is the transition coefficient for the refractor element that characterizes the propagation of light signal through the \gls{LC}-based \gls{RIS} elements.

The channel gain for the \gls{LoS} path can be given as~\cite{1277847}
\begin{equation} \label{eq: The channel gain of the LoS link}
    H_{\textnormal{LoS}}= \begin{cases} \frac{(m+1)A_{\textnormal{PD}}}{2\pi d^2} G(\xi) T(\xi) \cos^m(\Phi) \cos(\xi), \\
    \ \ \qquad \qquad 0 \leq \xi \leq \xi_{\textnormal{FoV}} \\
    0, \qquad \qquad \xi > \xi_{\textnormal{FoV}} \end{cases}
\end{equation}
\noindent where $m=-{\left(\log_2\left(\cos \left(\Phi_{1/2} \right)  \right)  \right)}^{-1}$ is the Lambertian order of emission with $\Phi_{1/2}$ as the semi-angle at half power of the \gls{LED}, $A_{\textnormal{PD}}$ is the detector physical area of the \gls{PD}, $d$ is the link distance, $\Phi$ denotes the radiation angle, $\xi$ is the incident angle, $T(\xi)$ is the gain of the optical filter, and $G(\xi)=f^2/\sin^2{\xi_{\textnormal{FoV}}}$ is the optical gain of the non-imaging concentrator with an internal refractive index $f$. In~\eqref{eq: The channel gain of the LoS link}, the cosine of $\xi$ captures the impact of random device orientation on the signal propagation and can be expressed as~\cite{8540452}
\begin{equation} \label{eq: Cos Xi}
\begin{split}
\cos(\xi) = & \big(\frac{x_a - x_u}{d}\big) \cos(\beta) \sin(\alpha) + \\ & \big(\frac{y_a - y_u}{d}\big) \sin(\beta) \sin(\alpha) + \\
 &  \big(\frac{z_a - z_u}{d}\big) \cos(\alpha), 
\end{split}
\end{equation}
where $\left(x_a,y_a,z_a \right)$ and $\left(x_u,y_u,z_u \right)$ are the position vectors for the locations of the \gls{AP} and the user, respectively, and $\alpha$ and $\beta$ denote the polar and azimuth angles of the receiver, respectively. It is assumed in this paper that the channel state information is known at the transmitter.

\subsubsection{Mirror Array-Based OSTAR-RIS-Aided VLC Channel}
\label{subSec: Mirror Array-Based OSTAR-RIS-Aided VLC Channel}

The channel gain of the \gls{NLoS} path considering signal propagation from the \gls{AP} to the $k$-th mirror having an area $d A_k$ and reflection from the $k$-th mirror to the $u$-th user in Room $1$ can be given as~\cite{9276478} 
\begin{equation} \label{eq: The channel gain of the NLoS link - RIS}
    H_{\textnormal{NLoS}}^{\textnormal{RIS}_k}= \begin{cases} \rho_{\textnormal{RIS}}\frac{(m+1)A_{\textnormal{PD}}}{2\pi^2 (d_k^a)^2 (d_u^k)^2} {dA}_k G(\xi) T(\xi) \cos^m(\Phi_k^a) \\ \times \ \cos(\xi_k^a) \cos(\Phi_u^k) \cos(\xi_u^k), \ 0 \leq \xi_u^k \leq \xi_{\textnormal{FoV}} \\
    0, \ \xi_u^k > \xi_{\textnormal{FoV}} \end{cases}
\end{equation}
where $\rho_{\textnormal{RIS}}$ denotes the reflectivity of the \gls{RIS}, $d_k^a$ and $d_k^u$ are the link distance for \gls{AP}-\gls{RIS} and  \gls{RIS}-user paths, respectively, $\Phi_k^a$ is the radiation angle for the path from the \gls{AP} to the $k$-th reflective surface, $\xi_k^a$ is the incident angle for the path from the \gls{AP} to $k$-th reflective surface, $\Phi_u^k$ is the angle of irradiance for the path from reflective surface $k$ towards user $u$, and $\xi_u^k$ is the incident angle of the reflected signal from surface $k$ to user $u$. Note that ${\cos }\left( {{\xi_k^a}} \right)$ can be easily calculated using~\eqref{eq: Cos Xi} to capture the effects of the random device orientation while $\cos(\Phi_u^k)$ can be expressed as~\cite{9910023}
\begin{equation} \label{eq: Cos Phi}
\begin{split}
\cos(\Phi_u^k)= & \big(\frac{x_k - x_u}{d_k^u}\big) \sin(\gamma) \cos(\omega) + \\
& \big(\frac{y_k - y_u}{d_k^u}\big) \cos(\gamma) \cos(\omega) + \\
& \big(\frac{z_k - z_u}{d_k^u}\big) \sin(\omega), 
\end{split}
\end{equation}
with $\left(x_k,y_k,z_k\right)$ being the position vector of the $k$-th reflecting element, and $\gamma$ and $\omega$ are the yaw and roll angles of the mirror array-based \gls{RIS}.

\subsubsection{Liquid Crystal-Based OSTAR-RIS-Aided VLC Channel}
\label{subSec: Liquid Crystal-Based OSTAR-RIS-Aided VLC Channel}

The channel gain of the \gls{NLoS} signal path for the signal propagation from the \gls{AP}, through the $n$-th refracting element, and to the $\hat{u}$-th user in Room $2$ is given as $H_{\textnormal{NLoS}}^{\textnormal{RIS}_n} \times \psi_{\textnormal{LC}}$, where 
\begin{equation} \label{eq: The channel gain of the NLoS link - STAR- RIS}
\begin{split}
    H_{\textnormal{NLoS}}^{\textnormal{RIS}_n}=& \begin{cases} \frac{(m+1)A_{\textnormal{PD}}}{2\pi^2 (d_n^a)^2 (d_{\hat{u}}^n)^2} {dA}_n G(\xi) T(\xi) \cos^m(\Phi_n^a)\\
    \times \ \cos(\xi_n^a) \cos(\Phi_{\hat{u}}^n) \cos(\xi_{\hat{u}}^n), \ 0 \leq \xi_{\hat{u}}^n \leq \xi_{\textnormal{FoV}} \\
    0, \ \xi_{\hat{u}}^n > \xi_{\textnormal{FoV}} \end{cases} \\
\textnormal{with} \ \  & \cos(\Phi_{\hat{u}}^n)= \big(\frac{x_n - x_{\hat{u}}}{d_n^{\hat{u}}}\big) \sin(\gamma) \cos(\omega) + \\
\qquad \qquad & \big(\frac{y_n - y_{\hat{u}}}{d_n^{\hat{u}}}\big) \cos(\gamma) \cos(\omega) + \big(\frac{z_n - z_{\hat{u}}}{d_n^{\hat{u}}}\big) \sin(\omega), 
\end{split}
\end{equation}
where $\left(x_n,y_n,z_n\right)$ is the location  of the $n$-th refracting element. On the other hand, the transition coefficient of any refracting element, $\psi_{\textnormal{LC}}$, can be given by~\cite{9910023}
\begin{equation} \label{eq: Transition Coefficient}
\psi_{\textnormal{LC}} = T_{\textnormal{ac}}(\xi_{\hat{u}}^n) \times T_{\textnormal{ca}}(\theta), 
\end{equation}
where $T_{\textnormal{ac}}(\xi_{\hat{u}}^n)$ and $T_{\textnormal{ca}}(\theta)$ represent the angular transmittance as the optical signal enters and exits the LC cell. Since none of the incident light signals get absorbed at the surface of the LC cell, the angular transmittance can be expressed in terms of the angular reflectance as $T_{\textnormal{ac}}(\xi_{\hat{u}}^n) =(1-R_{\textnormal{ac}}(\xi_{\hat{u}}^n))$ and $T_{\textnormal{ca}}(\theta)=(1-R_{\textnormal{ca}}(\theta))$. According to~\cite{9910023}, the angular reflectance can be derived as 

\begin{equation} \label{eq: angular reflectance ca}
\begin{split}
R_{\textnormal{ac}}(\xi_{\hat{u}}^n) = & \frac{1}{2} \Bigg(\frac{\eta^2 \cos(\xi_{\hat{u}}^n) - \sqrt{\eta^2 - \sin^2(\xi_{\hat{u}}^n)}}{\eta^2 \cos(\xi_{\hat{u}}^n) + \sqrt{\eta^2 - \sin^2(\xi_{\hat{u}}^n)}}\Bigg)^2 + \\
&  \frac{1}{2} \Bigg(\frac{ \cos(\xi_{\hat{u}}^n) - \sqrt{\eta^2 - \sin^2(\xi_{\hat{u}}^n)}}{ \cos(\xi_{\hat{u}}^n) + \sqrt{\eta^2 - \sin^2(\xi_{\hat{u}}^n)}}\Bigg)^2,
\end{split}
\end{equation}
\begin{equation} \label{eq: angular reflectance ac}
\begin{split}
R_{\textnormal{ca}}(\theta) = &  \frac{1}{2} \Bigg(\frac{ \cos(\theta) - \sqrt{\eta_1^2 - \sin^2(\theta)}}{ \cos(\theta) + \sqrt{\eta_1^2 - \sin^2(\theta)}}\Bigg)^2 + \\
& \frac{1}{2} \Bigg(\frac{\eta_1^2 \cos(\theta) - \sqrt{\eta_1^2 - \sin^2(\theta)}}{\eta_1^2 \cos(\theta) + \sqrt{\eta_1^2 - \sin^2(\theta)}}\Bigg)^2,
\end{split}
\end{equation}
where $\eta=\eta_c/\eta_a$ and $\eta_1=\eta_a/\eta_c$ are the relative refractive indices with $\eta_c$ and $\eta_a$ being the refractive indices of the LC cell and air, respectively. It can be observed from~\eqref{eq: Transition Coefficient} to~\eqref{eq: angular reflectance ac} that the transition coefficient $\psi_{\textnormal{LC}}$ can be optimized by tuning the refractive index of the LC cell $\eta_c$. According to~\cite{saleh2019fundamentals}, tuning the refractive index involves adjusting the tilt angle of the LC molecules (i.e., molecular orientation) by applying an external electric field. Mathematically, the relation between the tilt angle $\Xi$ of an \gls{LC} molecule and the refractive index of the \gls{LC} cell is given by 
\begin{equation} \label{eq: Relationship between the refractive index and the tilt angle}
\frac{1}{\eta_c^2 (\Xi)} = \frac{\cos^2(\Xi)}{\eta_e^2} + \frac{\sin^2(\Xi)}{\eta_o^2}, 
\end{equation}
where $\eta_o$ and $\eta_e$ denote the \gls{LC} cell's ordinary and extraordinary refractive indices, respectively. The tilt angle can also be expressed as a function of the externally applied voltage $V_{\textnormal{E}}$ as~\cite{9910023}
\begin{equation} \label{eq: tilt angle is controlled by an externally applied voltage}
    \Xi= \begin{cases} 
    0, \ V_{\textnormal{E}} \leq V_{\textnormal{TH}} \\ 
    \frac{\pi}{2} - 2 \tan^{-1} \Bigg[\textnormal{exp}\Bigg(\frac{V_\textnormal{TH} -V_{\textnormal{E}}}{V_0}\Bigg)\Bigg], \ V_{\textnormal{E}} > V_{\textnormal{TH}}
 \end{cases}
\end{equation}
where $V_\textnormal{TH}$ is the critical voltage threshold above which the tilting process begins. 

In addition to its light steering capability, LC-based RIS can be used to provide light amplification for the emerging signal through the process of stimulated emission. In this process, the incident photons interact with the excited molecules (via an external voltage) of the \gls{LC} cell. This causes the molecules to drop to a lower energy level, creating new identical photons. When an optical signal of power $P_{\textnormal {in}}$ impinges on an \gls{LC} cell with the transition coefficient $\psi_{\textnormal{LC}}$, the output signal power after undergoing light amplification in the presence of an external electric field can be calculated using the Beer's absorption law as~\cite{demtroder2014laser}
\begin{equation} \label{eq: The intensity of the output beam}
P_{\textnormal {out}} =P_{\textnormal {in}} \times \textnormal{exp}(\Gamma D) \times \psi_{\textnormal{LC}}, 
\end{equation}
where $\Gamma$ is the amplification gain coefficient, which is given as~\cite{marinova1999photorefractive}
\begin{equation} \label{eq: Amplification gain coefficient}
\Gamma = \frac{2 \pi \eta_c^3 }{\cos(\xi_{\hat{u}}^n) \lambda} r_{\textnormal{eff}} E, 
\end{equation}
$D$ denotes the depth of the LC cell, and the term $\textnormal{exp}(\Gamma D)$ represents the exponential increase of the incident signal power. In~\eqref{eq: Amplification gain coefficient}, $\lambda$ represents the optical signal's wavelength, $E$ [V/m] is the external electric field, and $r_{\textnormal {eff}}$ is the electro-optic coefficient. It can be observed from~\eqref{eq: Amplification gain coefficient} that the amplification capability of the LC-based RIS depends on the refractive index $\eta_c$, the wavelength $\lambda$, and the external electric field. However, the external electric field $E=V_{\textnormal{E}}/D$ and the refractive index of the LC-cell are related through~\eqref{eq: Relationship between the refractive index and the tilt angle} and~\eqref{eq: tilt angle is controlled by an externally applied voltage}. Specifically,
\begin{equation} \label{eq: Electric Field}
\resizebox{0.485\textwidth}{!}{$ V_{\textnormal{E}} = V_{\textnormal{TH}} - \textnormal{log} \Bigg( - \tan \Bigg[ \frac{\tan^{-1} \Bigg( \frac{\eta_o \sqrt{(\eta^2_e -\eta^2_o)(\eta^2_e - \eta^2_c)} }{\eta_c (\eta^2_e - \eta^2_o)}\Bigg) }{2} - \frac{\pi}{4}\Bigg] \Bigg). $} 
\end{equation}
\indent As a result, we focus on optimizing the refractive index to maximize the amplification gain coefficient for a given optical signal of wavelength $\lambda$.

\section{Adopted Multiple Access Schemes}
\label{Sec: Adopted Multiple Access Schemes}

This work investigates two multiple access schemes, namely, power-domain \gls{NOMA} and \gls{RSMA} to boost the sum rate of the proposed system. While implementing these schemes, the number of intended users is assumed to be $U$ and they have been sorted in order of their channel gain $H_1 \geq ... \geq H_u$ and $H_{\hat{u}} \geq ... \geq H_U$~\cite{maraqa2021achievable}. 

\subsection{Power-domain NOMA Scheme}
\label{subSec: Power-domain NOMA Scheme}

Following the power-domain \gls{NOMA} scheme~\cite{9154358}, using the superposition coding principle, the \gls{AP} sends a superposed signal to the intended users, which is given as~\cite{10183987}
\begin{equation} \label{eq: superposition coding}
    x^{\textnormal{NOMA}} = (\sum_{u=1}^{U} \sqrt{c_u P_\textnormal{S}} s_u) + I_\textnormal{DC},
\end{equation}
\noindent where $P_\textnormal{S}$ denotes the electrical transmission power of the signal ($P_\textnormal{S}$ is equal to $(\frac{p}{q})^2$~\cite{10183987}, here $p$ represents the optical power and $q$ symbolizes the electrical to optical conversion ratio), $s_u$ denotes the modulated message signal meant for the $u$-th user, $I_\textnormal{DC}$ is a constant bias current introduced to guarantee a positive instantaneous intensity~\cite{7792590}, and $c_u$ denotes the \gls{NOMA} power allocation ratio and is given by~\cite{10183987}
\begin{equation} \label{eq: allocated power ratio}
c_u =  \begin{cases}
       \mu^{\textnormal{NOMA}} (1-\mu^{\textnormal{NOMA}})^{u-1}, \ \textnormal{if} \ 1 \leq u<U \\
       (1-\mu^{\textnormal{NOMA}})^{u-1}, \ \textnormal{if} \ u=U
       \end{cases}
\end{equation}
\noindent where $\mu^{\textnormal{NOMA}}$ is a constant that falls within the range of $(0.5,1]$ and $\sum_{u=1}^U c_u~=~1$~\cite{shen2023secrecy}. After removing the DC bias, the signal received by the $u$-th user can be represented as~\cite{10183987}
\begin{equation} \label{eq: Received signal}
    y_u^{\textnormal{NOMA}} = H_u \times (\sum_{u=1}^{U} \sqrt{c_u P_\textnormal{S}} s_u) + z_u,
\end{equation} 
\noindent where $z_u \sim \mathcal{N}(0,\sigma^2)$ is the additive real-valued Gaussian noise with variance $\sigma^2$. Both the shot and thermal noises are included in $z_u$. According to the power-domain \gls{NOMA} scheme~\cite{9154358}, through \gls{SIC}, each user can decode its information. Subsequently, the sum rate of the proposed system under the power-domain \gls{NOMA} scheme is given by
\begin{equation} \label{eq: NOMA Sum rate}
\begin{aligned}
& R_{\textnormal{sum}}^{\textnormal{NOMA}} =\sum_{u=1}^{U} R_u, \\
& \textnormal{with} \\
& R_u = \begin{cases} B \textnormal{log}_2 \Bigg(1 + \frac{\textnormal{exp(1)}}{2 \pi} \frac{\big( R_{\textnormal{PD}} H_i \big)^2 c_i P_\textnormal{S}}{N_o B}  \Bigg), i=1 \\
B \textnormal{log}_2 \Bigg(1 + \frac{\textnormal{exp(1)}}{2 \pi} \frac{\big( R_{\textnormal{PD}} H_i \big)^2 c_i P_\textnormal{S}}{I_\textnormal{Room 1} + N_o B}  \Bigg), 1 < i \leq u\\
B \textnormal{log}_2 \Bigg(1 + \frac{\textnormal{exp(1)}}{2 \pi}  \frac{\big( R_{\textnormal{PD}} \textnormal{exp}(\Gamma D)  H_i  \big)^2 c_i P_\textnormal{S}}{I_\textnormal{Room 2} + N_o B}  \Bigg), \hat{u} \leq i \leq U
\end{cases}
\end{aligned}
\end{equation}
\noindent where $N_o$ represents the noise power spectral density, $B$ represents the system's bandwidth, both $I_\textnormal{Room 1}=(R_{\textnormal{PD}} H_i)^2 \sum_{j=1}^{i-1} c_j P_\textnormal{S}$ and $I_\textnormal{Room 2}=( R_{\textnormal{PD}} \textnormal{exp}(\Gamma D) H_i)^2 \sum_{j=1}^{i-1} c_j P_\textnormal{S}$ represent the inter-user interference terms in Room $1$ and Room $2$, respectively.

\subsection{RSMA Scheme}
\label{subSec: RSMA Scheme}
For implementing this scheme, the $1$-layer \gls{RSMA}~\cite{9461768,saeidi2023resource,9123680} is employed. In this scheme, a user's message is split into both a private part and a common part. The \gls{VLC} \gls{AP} combines all users' common parts into a single message, which is encoded into a common stream. This common stream is broadcast to all users alongside each user's private stream. Every user can decode the common stream, whereas each user only needs to decode their respective private stream. At the user-side, through the deployment of \gls{SIC}, the common stream is eliminated, allowing the user to decode their private stream, treating the private streams of other users as noise~\cite{saeidi2023resource}. Accordingly, the \gls{VLC} \gls{AP} transmitted power is divided as
\begin{equation} \label{eq: BS power}
    P_\textnormal{S} = P_\textnormal{0} + \sum_{u=1}^{U} P_u,
\end{equation} 
\noindent where $P_\textnormal{0}$ is the transmit power of the common stream and $P_u$ is the transmit power of the private stream intended for the $u$-th user. In this scheme, for a user to successfully decode its private stream, after subtracting the common stream, the difference between the common stream's power and the private streams' power should satisfy the following constraint~\cite{saeidi2023resource}
\begin{equation}\label{eq: SIC-RSMA}
    P_\textnormal{0} \delta_u - \sum_{u=1}^U P_u \delta_u \geq P_{\mathrm{tol}},
\end{equation}
\noindent where $\delta_u=\frac{|H_u|^2}{N_o}$ and $P_{\mathrm{tol}}$ is the \gls{SIC} threshold that guarantee the successful decoding of massages in \gls{RSMA} scheme. Based on~\eqref{eq: SIC-RSMA}, the \gls{AP} allocates a portion of its total power to the common stream (i.e., $P_\textnormal{0}=\mu^{\textnormal{RSMA}} P_\textnormal{S}$) and the other portion to the private streams (i.e., $\sum_{u=1}^U P_u = (1-\mu^{\textnormal{RSMA}}) P_\textnormal{S}$), where $\mu^{\textnormal{RSMA}} \in (0,1)$~\cite{alazwary2023rate}. The aforementioned portion of power dedicated to private streams is divided equally between users. In the simulation section, we show that the adopted power allocation (i.e., equal power allocation) strategy is better performance-wise than both (i) the \gls{NOMA}-alike power allocation strategy (i.e., distributing the power using the same strategy utilized in the \gls{NOMA} scheme based on~\eqref{eq: allocated power ratio}) and (ii) the random power allocation. The \gls{VLC} \gls{AP} transmitted signal is given as~\cite{9461768} 
\begin{equation} \label{eq: transmit signal RSMA}
    x^\textnormal{RSMA} = \sqrt{P_\textnormal{0}} s_c + \sum_{u=1}^{U} \sqrt{P_u} s_u + I_\textnormal{DC},
\end{equation} 
\noindent where $s_c$ denotes the common stream, $s_u$ denotes the private stream intended for the $u$-th user, and $I_\textnormal{DC}$ is a constant bias current. At the $u$-th user, the total received signal is 
\begin{equation} \label{eq: received signal RSMA}
    y_u^\textnormal{RSMA} = H_u \times (\sqrt{P_\textnormal{0}} s_c + \sum_{u=1}^{U} \sqrt{P_u} s_u) + z_u.
\end{equation} 
\indent Accordingly, the rate of the $u$-th user decoding the common stream is given as
\begin{equation} \label{eq: rate of user u decoding common stream}
\begin{aligned}
& R_{c,u} = \begin{cases} B \textnormal{log}_2 \Bigg(1 + \frac{\textnormal{exp(1)}}{2 \pi} \frac{\big( R_{\textnormal{PD}} H_i \big)^2 P_\textnormal{0}}{ I_\textnormal{Room 1}^\textnormal{Common} + N_o B}  \Bigg), 1 \leq i< u \\
B \textnormal{log}_2 \Bigg(1 + \frac{\textnormal{exp(1)}}{2 \pi} \frac{\big( R_{\textnormal{PD}} \textnormal{exp}(\Gamma D) H_i \big)^2 P_\textnormal{0}}{ I_\textnormal{Room 2}^\textnormal{Common} + N_o B}  \Bigg), \hat{u} \leq i \leq U
\end{cases}
\end{aligned}
\end{equation}
\noindent where both $I_\textnormal{Room 1}^\textnormal{Common}=( R_{\textnormal{PD}} H_i)^2 \sum_{j=1}^U P_j$ and $I_\textnormal{Room 2}^\textnormal{Common}=(R_{\textnormal{PD}} \textnormal{exp}(\Gamma D) H_i)^2 \sum_{j=1}^U P_j$ represent an inter-user interference terms, resulting from the application of the \gls{RSMA} scheme, in Room $1$ and Room $2$, respectively.

To ensure the successful decoding of the common stream by all users, based on the assumed channel ordering, we should choose the rate of the common stream to be $R_{c,U}$~\cite{9461768}, which can be represented as
\begin{equation} \label{eq: rate of common stream}
\begin{split}
\underset{1 \leq u \leq U}{\text{min}} R_{c,u} =&  R_{c,U} \\ =& B \textnormal{log}_2 \Bigg(1 + \frac{\textnormal{exp(1)}}{2 \pi} \times \\ & \resizebox{0.3\textwidth}{!}{$ \frac{\big( R_{\textnormal{PD}} \textnormal{exp}(\Gamma D) H_U \big)^2 P_\textnormal{0}}{\big( R_{\textnormal{PD}} \textnormal{exp}(\Gamma D) H_U \big)^2 \sum_{j=1}^U P_j + N_o B}  \Bigg).$}
\end{split}
\end{equation} 
\indent Given that the rate of the common stream is $R_{c,U}$, the total data rates of all users receiving the common stream must be less than or equal $R_{c,U}$. This relation can be expressed as~\cite{9461768}
\begin{equation} \label{eq: common rate constraint}
    \sum_{u=1}^U R_{c,u} \leq R_{c,U}, 
\end{equation} 
\noindent where, based on Lemma $1$ in~\cite{9461768}, the optimal solution for the rate of the common stream is obtained at the equality of~\eqref{eq: common rate constraint}.

At each user, the rate of decoding the private stream is
\begin{equation} \label{eq: achievable rate of user u decoding private stream}
\begin{aligned}
& R_{p,u} = \begin{cases} B \textnormal{log}_2 \Bigg(1 + \frac{\textnormal{exp(1)}}{2 \pi} \frac{\big( R_{\textnormal{PD}} H_i \big)^2 P_i}{ I_\textnormal{Room 1}^\textnormal{Private} + N_o B}  \Bigg), 1 \leq i< u \\
\resizebox{0.405\textwidth}{!}{$ B \textnormal{log}_2 \Bigg(1 + \frac{\textnormal{exp(1)}}{2 \pi} \frac{\Big( R_{\textnormal{PD}} \textnormal{exp}(\Gamma D) H_i \Big)^2 P_i}{ I_\textnormal{Room 2}^\textnormal{Private} + N_o B}  \Bigg), \hat{u} \leq i \leq U $}
\end{cases}
\end{aligned}
\end{equation}
\noindent where both $I_\textnormal{Room 1}^\textnormal{Private}=( R_{\textnormal{PD}} H_i )^2 \sum_{j=1, \ j \neq u}^U P_j$ and $I_\textnormal{Room 2}^\textnormal{Private}=(R_{\textnormal{PD}} \textnormal{exp}(\Gamma D) H_i)^2 \sum_{j=1, \ j \neq u}^U P_j$ represent also an inter-user interference terms in Room $1$ and Room $2$, respectively.

Subsequently, the sum rate of the proposed \gls{RSMA} scheme is given by
\begin{equation} \label{eq: RSMA sum rate}
    R_{\textnormal{sum}}^{\textnormal{RSMA}} =\sum_{u=1}^U (R_{c,u} + R_{p,u}). 
\end{equation}

\section{Sum Rate Optimization}
\label{Sec: Sum Rate Optimization}
In this section, the details of the sum rate optimization problem for both power-domain \gls{NOMA} and \gls{RSMA} schemes is presented. Also, the sine-cosine meta-heuristic algorithm is utilized to get the global optimal solution for the formulated multi-variate non-convex optimization problem.

\subsection{Sum Rate Maximization Problem}
The main purpose of this paper is to maximize the sum rate of the \gls{OSTAR-RIS}-aided VLC system. To do so, the sum rate maximization problem that jointly optimizes the roll angle, $\omega$, the yaw angle, $\gamma$, and the refractive index, $\eta_c$, of the \gls{OSTAR-RIS} is considered and formulated as 
\begin{alignat}{3}
\textnormal{(P0):} &\underset{\{\omega,\gamma,\eta_c\}}{\text{max}} &\ & R_{\textnormal{sum}}^{\textnormal{NOMA}} \ \textnormal{or} \ R_{\textnormal{sum}}^{\textnormal{RSMA}}, \label{eq:objective1}\\
&\quad \ \text{s.t.} &  & - \frac{\pi}{2} \leq \omega \leq \frac{\pi}{2}, \label{eq:constraint-a} \\
&  &  & - \frac{\pi}{2} \leq \gamma \leq \frac{\pi}{2}, \label{eq:constraint-b}\\ 
&  &  &  \quad 1.5 \leq \eta_c \leq 1.7, \label{eq:constraint-c} 
\end{alignat}
\noindent where the constraints~\eqref{eq:constraint-a} and~\eqref{eq:constraint-b} denote the bounds of the roll angle and yaw angle, respectively, of the mirror array-based \gls{OSTAR-RIS} elements. Constraint~\eqref{eq:constraint-c} denotes the bound of the \gls{LC}-based \gls{OSTAR-RIS} elements. (P0) requires the joint optimization of all three optimization variables. Besides, it is highly non-convex. Subsequently, it cannot be solved using traditional optimization methods. To get the global optimal solution of (P0), we resort to the \gls{SCA}~\cite{mirjalili2016sca}, which is a population-based meta-heuristic algorithm. The merits of the \gls{SCA} method over other metaheuristics like ant colony optimization, particle swarm optimization, and genetic algorithm are as follows, (i) ability to avoid being trapped in a local maxima (i.e., a sub-optimal solution), (ii) fast convergence, (iii) ease of implementation, and (iv) fewer parameters~\cite{mirjalili2016sca}.

\subsection{Proposed Solution Methodology}
\label{subsection: Proposed Solution Approach}

The \gls{SCA}, a method based on iterative population search, was developed by Mirjalili~\cite{mirjalili2016sca} and has recently found applications in wireless communications~\cite{9543660,9910023,10183987,10114573,10202576}, remote sensing~\cite{9408235}, and electrical power system design~\cite{9715094,8598859} due to its ability to obtain the global optimal solution for complex optimization problems. According to this method, $G$ search agents are randomly deployed within the boundary of the search space of the optimization problem. The position of each agent represents a candidate solution to the optimization problem and gets updated via mathematical models based on the sine and cosine functions. Thus, the agents explore the search space in parallel to find the promising regions by using the sine function and then exploit those regions to obtain the overall best solution via the cosine function. At iteration $t$, the solution of agent $g$ is denoted as ${\bf s}_{g}^t= [ \omega, \gamma, \eta_c]$ and its fitness can be determined via the objective function \eqref{eq:objective1}. The fitness of the agents is evaluated using the objection function and the current best solution among them at iteration $t$ (i.e., fittest agent), called destination point, is denoted as ${\mathcal D}^t$. It is called a destination point since the other agents would try to get to that position in the next iteration. At the $t+1$-th iteration, the solution of the $v$-th variable of each agent is updated as follows~\cite{mirjalili2016sca}
\begin{equation} \label{eq: current solution}
    s_{g,\upsilon}^{t+1}= \begin{cases} 
     s_{g,\upsilon}^{t} + r_1 \times \cos (r_2) \times |r_3 \mathcal{D}^t - s_{g,\upsilon}^{t} |, \ \textnormal{if} \ r_4 \geq 0.5 \\
    s_{g,\upsilon}^{t} + r_1 \times \sin (r_2) \times |r_3 \mathcal{D}^t - s_{g,\upsilon}^{t} |, \ \textnormal{if} \ r_4 < 0.5
 \end{cases}
\end{equation}
where $r_1, r_2,r_3$ and $r_4$ are the main parameters that influence the search procedure of the \gls{SCA} and $|\cdot|$ denotes the absolute value. Specifically
\begin{equation} \label{eq: update r1}
r_1={\widetilde{a}}-t({\widetilde{a}}/T), 
\end{equation}
with ${\widetilde{a}}$ being a constant and $T$ being the maximum number of iterations, can adaptively guide the movement direction of the search agents (i.e., the location to be searched in the next iteration). The global exploitation and local exploration abilities of the \gls{SCA} are emphasized when $r_1 > 1$ and $r_1 \le 1$, respectively. As evident from~\eqref{eq: update r1}, the value of $r_1$ is larger during the early stage of the algorithm (since $t$ is small), which enables excellent global exploitation. As $t$ gets larger, the value of $r_1$ decreases, which allows the algorithm to thoroughly search (i.e., local exploration) around the current best solution for the global optimal solution. The parameter  $r_2$ controls the movement distance towards or outwards the destination point and its value is selected randomly from the interval $(0,2\pi)$. The parameter $r_3$ determines the extent to which the destination point affects the distance between the current solution and the destination point and its value is randomly chosen according to a uniform distribution in the interval $(0,2)$. Lastly, the parameter $r_4$ which takes on a random value in the interval $(0,1)$ switches with equal probability between the sine and cosine functions. The components $r_1 \times \sin(r_2)$ and $r_1 \times \cos(r_2)$ jointly enhance the exploration (when these two values are greater than 1 or less than -1) and exploitation (when these two values fall within the range of -1 to 1) during the search process. In summary, the parameters $r_1$, $r_2$, $r_3$, and $r_4$ are used to iteratively guide the movement of the agents in the search space to achieve a good balance between exploration (covering a wide range of solutions) and exploitation (focusing on promising solutions). For each iterate, the fitness of all agents is determined and the fittest serves as the new destination point. This repeats until a predefined termination criterion is met. Algorithm~\ref{Algorithm: proposed solution 1} summarizes the proposed \gls{SCA}-based algorithm.

\SetAlgoNlRelativeSize{-1}
\begin{algorithm}[!t]
\scriptsize
\DontPrintSemicolon
\SetAlgoLined
\caption{Proposed solution for optimization problem (P0).} \label{Algorithm: proposed solution 1}
\LinesNumbered
\KwIn{$G$, $T$, ${\widetilde{a}}$, $V$, $V_\textnormal{TH}$, $V_\textnormal{E}$, $V_\textnormal{0}$, $\rho_\textnormal{RIS}$, $\eta_e$, $\eta_a$, $\eta_o$;}
\KwOut{The optimal yaw and roll angles and the refractive index $\gamma^*$, $\omega^*$, and 
 $\eta_c^*$, respectively;}
Initialize $t=0$ and $s_{g,\upsilon}^{t}$, $\forall g$;\;
Compute the fitness of each agent via~\eqref{eq: NOMA Sum rate} or~\eqref{eq: RSMA sum rate};\;
Set $\mathcal{D}^t$ as the solution of the fittest agent;\;
Update $t=t+1$;\;
\While{No convergence}{
Obtain $r_1$,$r_2$, $r_3$, $r_4$;\;
\For{$g=1:G$}{
 \For{$\upsilon=1:3$}{ 
 Update $s_{g,\upsilon}^{t}$, $\forall g$ by solving~\eqref{eq: current solution};\;
}
}
Check agents for constraint(s) violation;\; 
Update the fitness of each agent using~\eqref{eq: NOMA Sum rate} or~\eqref{eq: RSMA sum rate};\;
Update~$\mathcal{D}^t$;\;
Update $t=t+1$;\;
}
\end{algorithm}

\subsection{Computational Complexity Analysis}
\label{Subsection: Computational Complexity Analysis}
The time complexity of Algorithm~\ref{Algorithm: proposed solution 1} mainly depends on (i) the generation of the initial set of solutions for all agents, (ii) the evaluation of the fitness of the solution of all agents, (iii) the selection of the destination point, (iv) updating the agents' solutions, and (v) the total number of iterations. The corresponding time complexities of the tasks mentioned in (i), (ii), (iii), and (iv) are ${\mathcal O} (GV)$, ${\mathcal O} (G)$, ${\mathcal O} (G)$,  and ${\mathcal O} (GVT)$, respectively, where $V$ denotes the total number of decision variables. As a result, the worst-case computational complexity of Algorithm~\ref{Algorithm: proposed solution 1} is  ${\mathcal O} (GVT)$.

\section{Sum Energy Efficiency Optimization}
\label{Sec: Sum Energy Efficiency Optimization}

The \gls{SEE} metric has gained widespread use as a performance metric in \gls{VLC} systems~\cite{10183987,9350598,8307185}. It can be defined as the ``ratio between the sum rate of the \gls{VLC} system and the total power the system consumes'', and is typically measured in Bits per Joule. The sum rate of the proposed system, for the power-domain \gls{NOMA} and the \gls{RSMA} schemes, are detailed in~\eqref{eq: NOMA Sum rate} and~\eqref{eq: RSMA sum rate}, respectively. The total consumed power of the proposed system encompasses various components: Firstly, the consumed power at the \gls{VLC} \gls{AP} can be expressed as~\cite{10183987}
\begin{equation}
\mathcal{P}_\textnormal{VLC-AP} =  \mathcal{P}_\textnormal{T-Circuit}+ \mathcal{P}_\textnormal{Driver} + \mathcal{P}_\textnormal{PA} + \mathcal{P}_\textnormal{Filter} + \mathcal{P}_\textnormal{DAC} + \mathcal{P}_\textnormal{S},    
\end{equation}
\noindent where $\mathcal{P}_\textnormal{T-Circuit}$ denotes the power of the transmitter external circuit, $\mathcal{P}_\textnormal{Driver}$ denotes the power of the \gls{LED} driver, $\mathcal{P}_\textnormal{PA}$ denotes the power of the power amplifier, $\mathcal{P}_\textnormal{Filter}$ denotes the power of the filter, $\mathcal{P}_\textnormal{DAC}$ denotes the power of the digital-to-analog converter (DAC), and $\mathcal{P}_\textnormal{S}$ denotes the power of the signal. Secondly, the power consumed at the \gls{OSTAR-RIS} can be expressed as
\begin{equation}
\mathcal{P}_\textnormal{OSTAR-RIS} =  \mathcal{P}_\textnormal{m} \times \mathcal{K} + \mathcal{P}_\textnormal{LC} \times \mathcal{N},   
\end{equation}
\noindent where $\mathcal{P}_\textnormal{m}$ denotes the power of each mirror array-based reflector element, and $\mathcal{P}_\textnormal{LC}$ denotes the power of each \gls{LC}-based refractor element. Lastly, the power consumption at the receiver can be expressed as~\cite{10183987}
\begin{equation}
\mathcal{P}_\textnormal{R} = \mathcal{P}_\textnormal{R-Circuit}+ \mathcal{P}_\textnormal{Filter} + \mathcal{P}_\textnormal{TIA} + \mathcal{P}_\textnormal{ADC},
\end{equation}
\noindent where $\mathcal{P}_\textnormal{R-Circuit}$ denotes the power of the receiver external circuit, $\mathcal{P}_\textnormal{TIA}$ denotes the power of the trans-impedance amplifier (TIA), and $\mathcal{P}_\textnormal{ADC}$ denotes the power of the analog-to-digital converter (ADC). Overall, the total consumed power of the proposed system can be given as  
\begin{equation} \label{eq: total power consumption}
\mathcal{P}_{\textnormal{Total}} =  \mathcal{P}_\textnormal{VLC-AP} + \mathcal{P}_\textnormal{OSTAR-RIS} + \mathcal{P}_\textnormal{R}.  
\end{equation}
\indent Consequently, the \gls{SEE} of the proposed system, for the power-domain \gls{NOMA} and the \gls{RSMA} schemes, can be formulated, respectively, as
\begin{equation} \label{eq: SEE}
\begin{split}
    \textnormal{SEE}^\textnormal{NOMA} = & \frac{R_{\textnormal{sum}}^{\textnormal{NOMA}}}{\mathcal{P}_{\textnormal{Total}}}, \\     \textnormal{SEE}^\textnormal{RSMA} = & \frac{R_{\textnormal{sum}}^{\textnormal{RSMA}}}{\mathcal{P}_{\textnormal{Total}}}.
\end{split}
\end{equation}
\indent To maximize the \gls{SEE} metric, the objective function of (P0) is replaced by~\eqref{eq: SEE}. Then, the resultant optimization problem is solved using similar procedures to the ones outlined in Section~IV-B.

\begin{table}[!t]
\centering
\vspace{-2em}
\caption{Simulation Parameters}
\vspace{-0.5em}
\label{tab: Simulation Parameters}
\resizebox{\columnwidth}{!}{%
\begin{tabular}{|l|l||l|l|l}
\cline{1-4}
\textbf{Parameter} & \textbf{Value} & \textbf{Parameter} & \textbf{Value} &  \\ \cline{1-4}

$\Phi_{1/2}$ & $ 70^\circ$ & $A_{\textnormal{PD}}$ & $1.0 \ \textnormal{cm}^2$ &  \\ \cline{1-4}

$T(\xi)$ & $1.0$ & $d$ & $2.5 \ \textnormal{m} $  & \\ \cline{1-4}
 
$f$ & $1.5$ & $\xi_{\textnormal{FoV}}$ & $85^{\circ}$ & \\ \cline{1-4}
 
$\rho_\textnormal{RIS}$ & $0.95$ & $\rho_\textnormal{wall}$ & $0.8$ & \\ \cline{1-4}
 
$\eta_e$ & $1.7$ & $\eta_a$ & $1.0$ &  \\ \cline{1-4}

$V_\textnormal{TH}$ & $1.34$ V & $\eta_o$ & $1.5$ & \\ \cline{1-4}

$D$ & $0.75$ mm & $V_0$ & $1.0$ V & \\ \cline{1-4}

$r_\textnormal{eff}$ & $12$ pm/V & $\lambda$ & $\{510,670\}$ nm & \\ \cline{1-4}

$q$ & $3.0$ & $B$ & $200$ MHz & \\ \cline{1-4}
 
$N_o$ & $10^{-21} \textnormal{A}^2/\textnormal{Hz}$ & $R_\textnormal{PD}$ & $0.53$ A/W &  \\ \cline{1-4}

$V$ & $3$ & $G$ & $5$ &  \\ \cline{1-4}

$T$ & $4000$ & ${\widetilde{a}}$ & $2.0$ &  \\ \cline{1-4}

$U$ & $4$ & $\{\mu^{\textnormal{NOMA}},\mu^{\textnormal{RSMA}}\}$ & $\{0.6,0.6\}$ & \\ \cline{1-4}

$P_{\mathrm{tol}}$ & $10$ dBm & $p$ & $[1,1.5,...,4]$ Watt &  \\ \cline{1-4}

$\mathcal{P}_\textnormal{ADC}$ & $95$ mWatt & $\mathcal{P}_\textnormal{DAC}$ & $175$ mWatt & \\ \cline{1-4}

$\mathcal{P}_\textnormal{Driver}$ & $2758$ mWatt & $\mathcal{P}_\textnormal{Filter}$ & $2.5$ mWatt & \\ \cline{1-4}

$\mathcal{P}_\textnormal{m}$ & $100$ mWatt & $\mathcal{P}_\textnormal{TIA}$ & $2500$ mWatt & \\ \cline{1-4}

$\mathcal{P}_\textnormal{R-Circuit}$ & $1.9$ mWatt & $\mathcal{P}_\textnormal{T-Circuit}$ & $3250$ mWatt & \\ \cline{1-4}

$\mathcal{P}_\textnormal{LC}$ & $320$ mWatt & $\mathcal{P}_\textnormal{PA}$ & $280$ mWatt &\\ \cline{1-4}

\end{tabular}%
}
\end{table}

\section{Simulation Results}
\label{Sec: Simulation Results}
This section presents extensive numerical simulations to assess the performance of the proposed \gls{OSTAR-RIS}-aided \gls{VLC} system under both the power-domain \gls{NOMA} and the \gls{RSMA} schemes. Table~\ref{tab: Simulation Parameters} summarizes the default parameters used during the simulations. Beyond this list, the polar angle, $\alpha$, and the azimuth angle, $\beta$, are characterized using the uniform distribution within the range of $[-\pi,\pi]$ and the Laplace distribution within the range of $[0,\frac{\pi}{2}]$, respectively. The system depicts users as cylindrical objects with a height of $1.65$ meters and a diameter of $0.3$ meters, with a receiver that is located $0.36$ meters away from their body at a height of $0.85$ meters. Both Rooms $1$ and $2$ have a dimension of $5.0 \times 5.0 \times 3.0$ meters with an \gls{AP} fixed at the center of Room $1$. The \gls{OSTAR-RIS} comprises $5$ rows and $10$ columns with $25$ \gls{LC}-based elements and $25$ mirror array-based elements, unless otherwise specified. Each element is a square that spans over $0.1~\textnormal{m} \times 0.1~\textnormal{m}$. In this section, we set $\iota$ in~\eqref{eq: Total Channel Gain} to zero, as we focus on the \gls{NLoS} scenario. The parameters utilized in this section are extracted from~\cite{maraqa2021achievable, 9910023, 9543660, 8307185,zhan2022optimal,10183987,saeidi2023resource,Si-PIN-photodiode}. 

\begin{figure}[!t]
\centering
\includegraphics[width=0.485\textwidth]{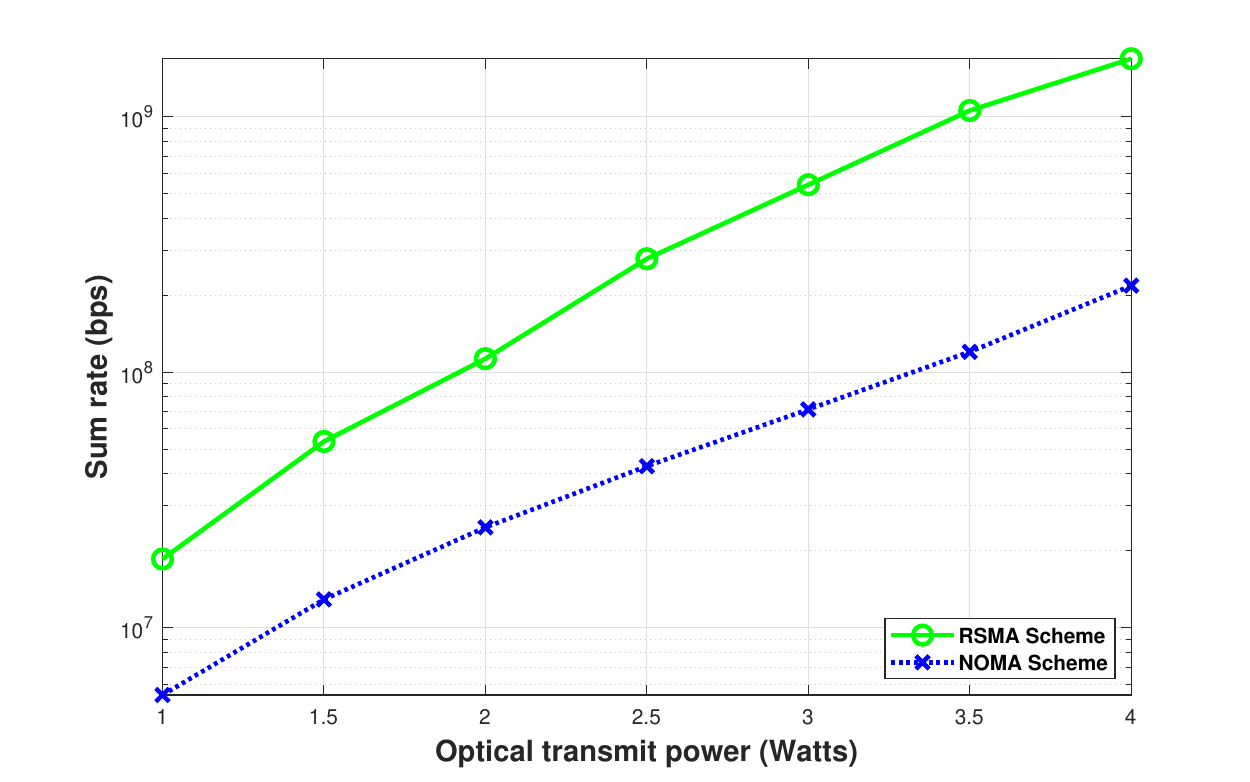}
\caption{Sum rate versus \gls{AP} optical transmit power. At $\lambda = 510$ nm.}
\label{fig: A_Tx_Power_vs_Rate}
\vspace{-1em}
\end{figure}

\begin{figure}[!t]
\centering
\includegraphics[width=0.485\textwidth]{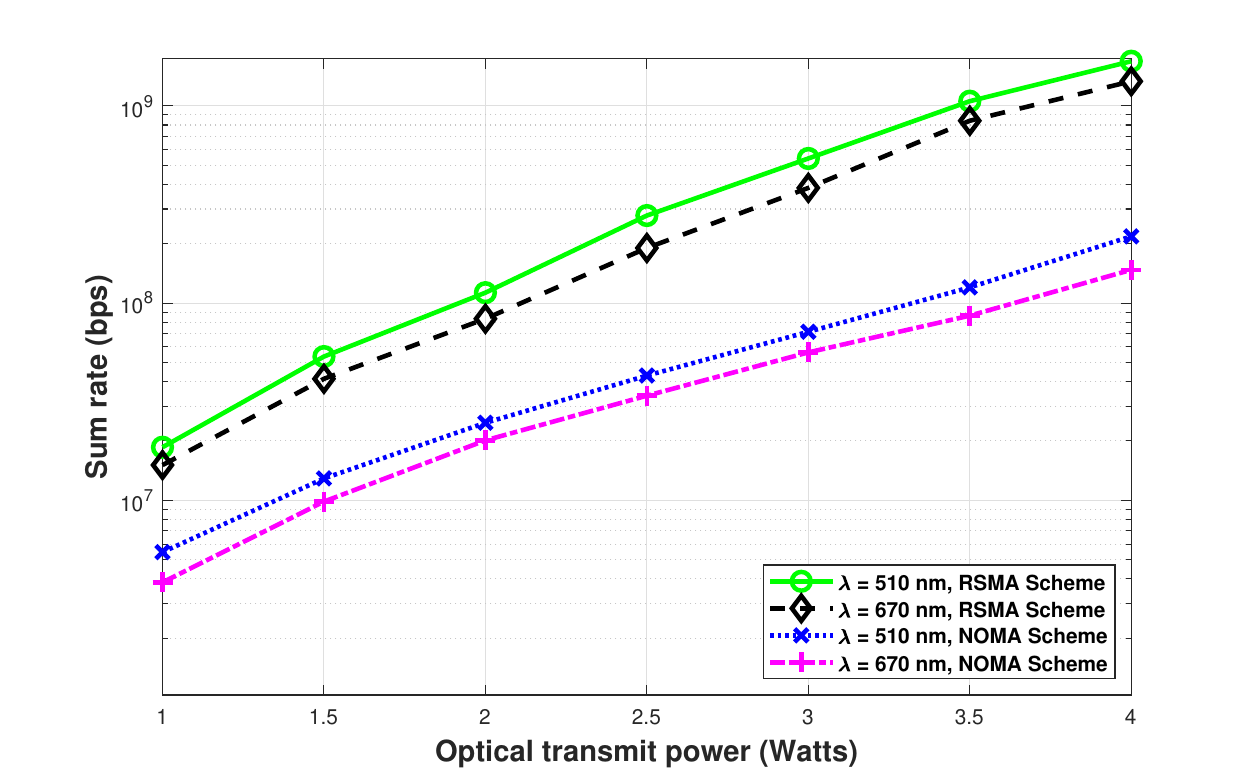}
\caption{Sum rate versus wavelength of the transmitted light signal. At $p = 3$ Watts.}
\label{fig: C_P_vs_R_wavelenght}
\vspace{-1.5em}
\end{figure}

\begin{figure}[!t]
\centering
\includegraphics[width=0.485\textwidth]{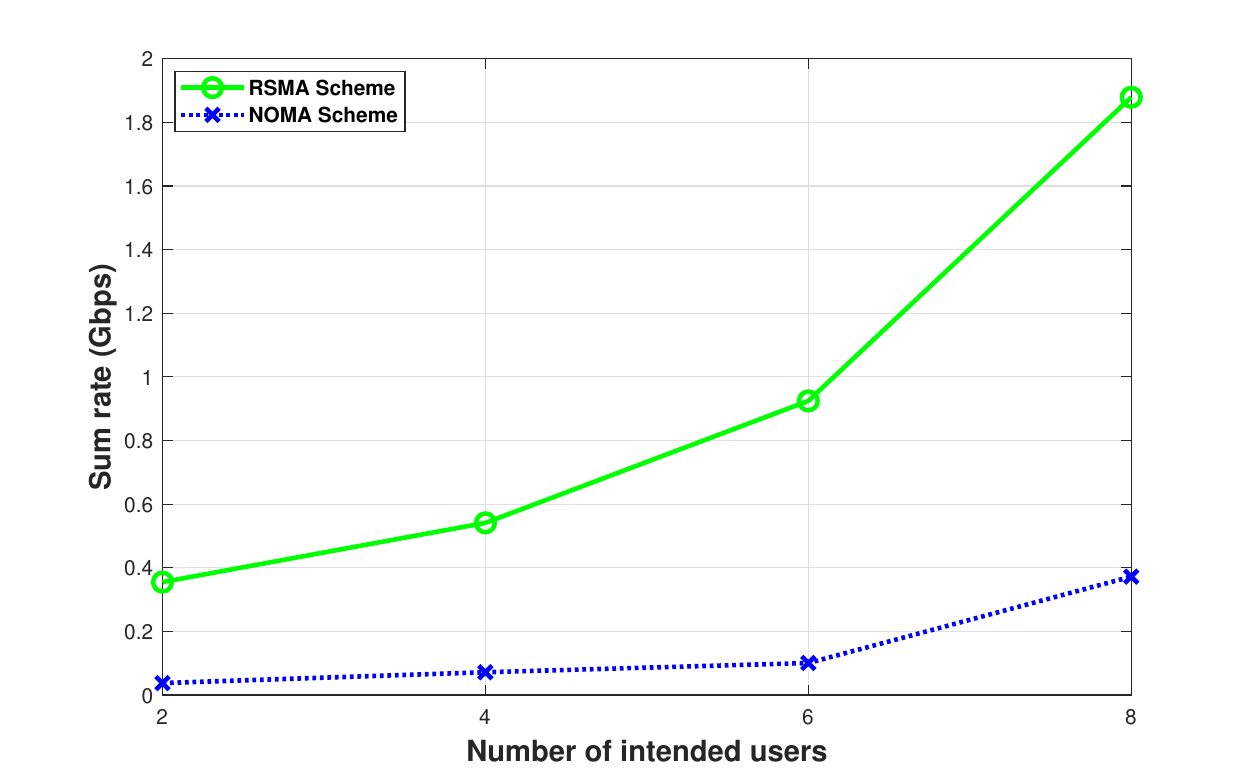}
\caption{Sum rate versus number of intended users. At $\lambda = 510$ nm, and $p = 3$ Watts.}
\label{fig: F_Variable_users}
\vspace{-1em}
\end{figure}

Fig.~\ref{fig: A_Tx_Power_vs_Rate} compares the sum rate performance of the proposed \gls{OSTAR-RIS}-assisted multi-user indoor \gls{VLC} system for the \gls{RSMA} scheme and the power-domain \gls{NOMA} scheme for different \gls{AP} optical transmit powers. It can be observed that the sum-rate performance of \gls{RSMA}-aided system is superior to the corresponding \gls{NOMA}-aided system with an enhancement in the sum rate that is up to $456$\%. This illustrates that adopting the \gls{RSMA} scheme is more spectrally efficient for the proposed \gls{OSTAR-RIS}-assisted \gls{VLC} system (i.e., achieves higher date performance with the same resources) than the power-domain \gls{NOMA} scheme.

Fig.~\ref{fig: C_P_vs_R_wavelenght} compares the sum rate performance of the proposed \gls{OSTAR-RIS}-assisted multi-user indoor \gls{VLC} system for the \gls{RSMA} scheme and the power-domain \gls{NOMA} scheme for different wavelength of the transmitted light signal. This figure is generated because altering the wavelengths affects the sum-rate performance of the \gls{LC}-based \gls{OSTAR-RIS} elements~\cite{9910023,10183987} (i.e., based on~\eqref{eq: Amplification gain coefficient}, as the wavelength increases, the amplification gain coefficient, $\Gamma$, decreases and subsequently, the sum rate performance deteriorates). This figure shows a sum-rate deterioration up to $31$\% and $29$\% in \gls{RSMA}-aided system and power-domain \gls{NOMA}-aided system, respectively, when the wavelength of the transmitted light signal is changed from $510$ nm to $670$ nm. 

\begin{figure}[!t]
\centering
\vspace{-1em}
\includegraphics[width=0.485\textwidth]{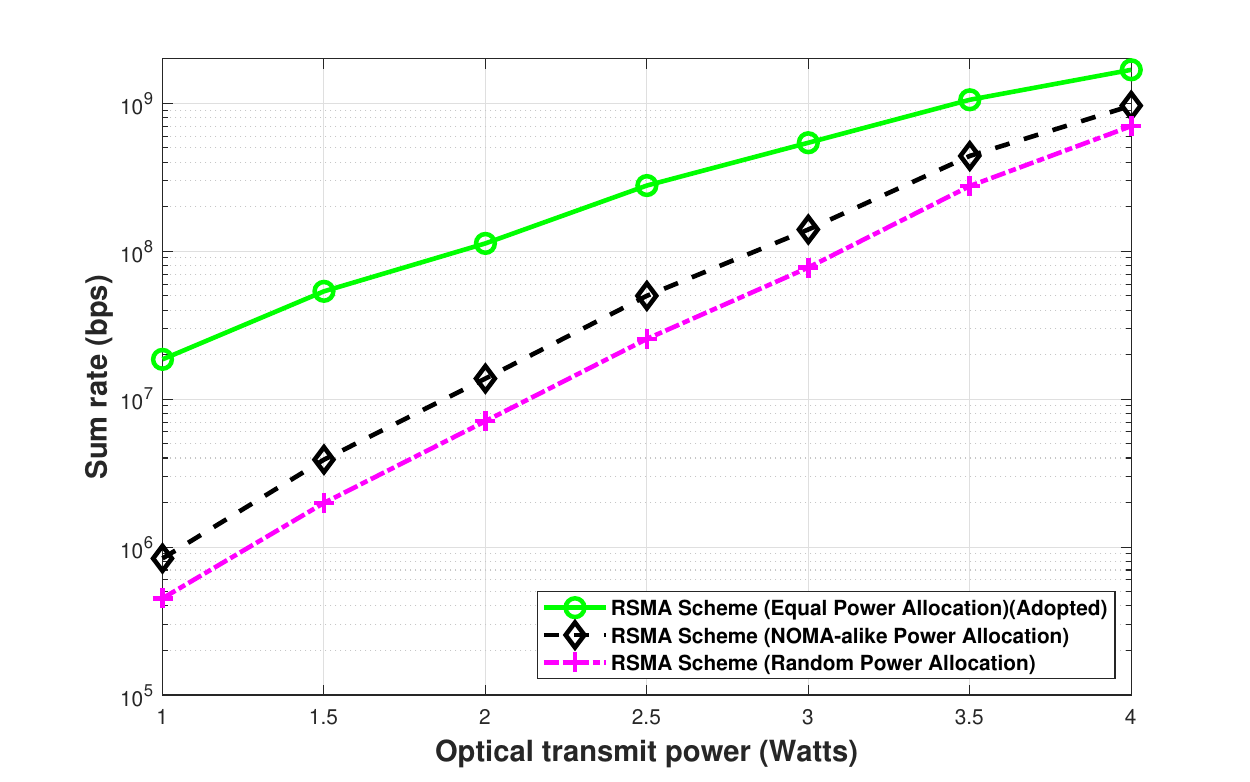}
\caption{Sum rate versus \gls{AP} optical transmit power for different power allocation strategies for the \gls{RSMA} scheme. At $\lambda = 510$ nm and $p = 3$ Watts.}
\label{fig: E_Fixed_NOMAalike_Random}
\vspace{-1.5em}
\end{figure}

\begin{figure}[!t]
    \centering
    \vspace*{-0.2in}
    \subfloat[Sum rate.]{
        \hspace*{-2em}
        \includegraphics[scale=0.4]{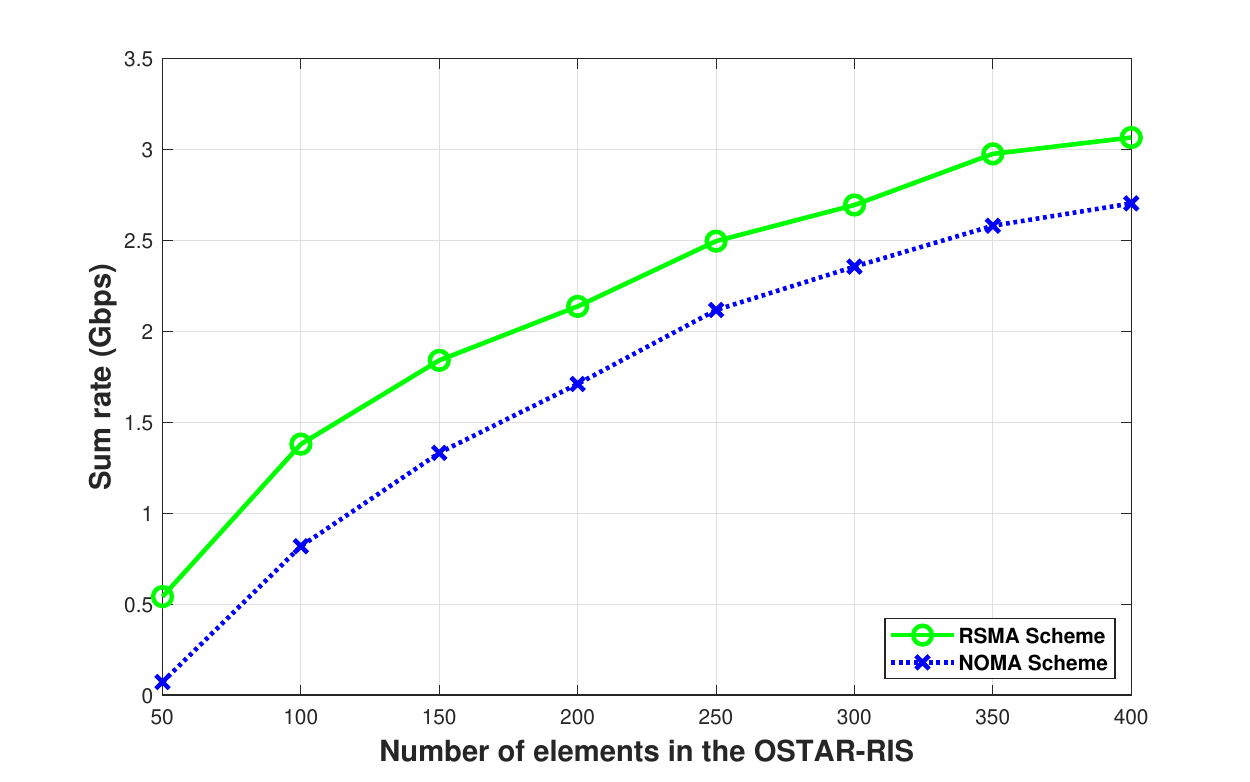}
        \label{fig: D_mirrors_vs_R}
    }
    \hfill
    \subfloat[Total consumed power.]{
        \hspace*{-2em}
        \includegraphics[scale=0.4]{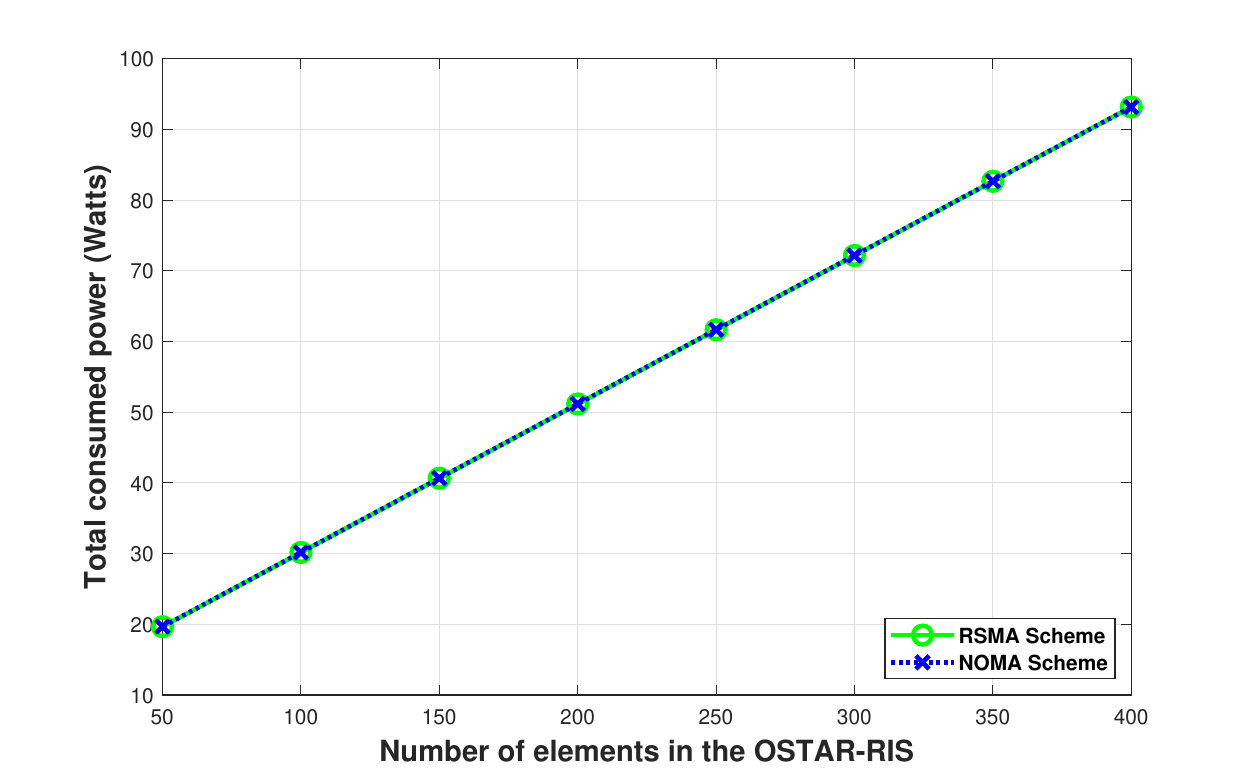}
        \label{fig: D_mirrors_vs_P}
    }
    \hfill
    \subfloat[Sum energy efficiency.]{
        \hspace*{-2em}
        \includegraphics[scale=0.4]{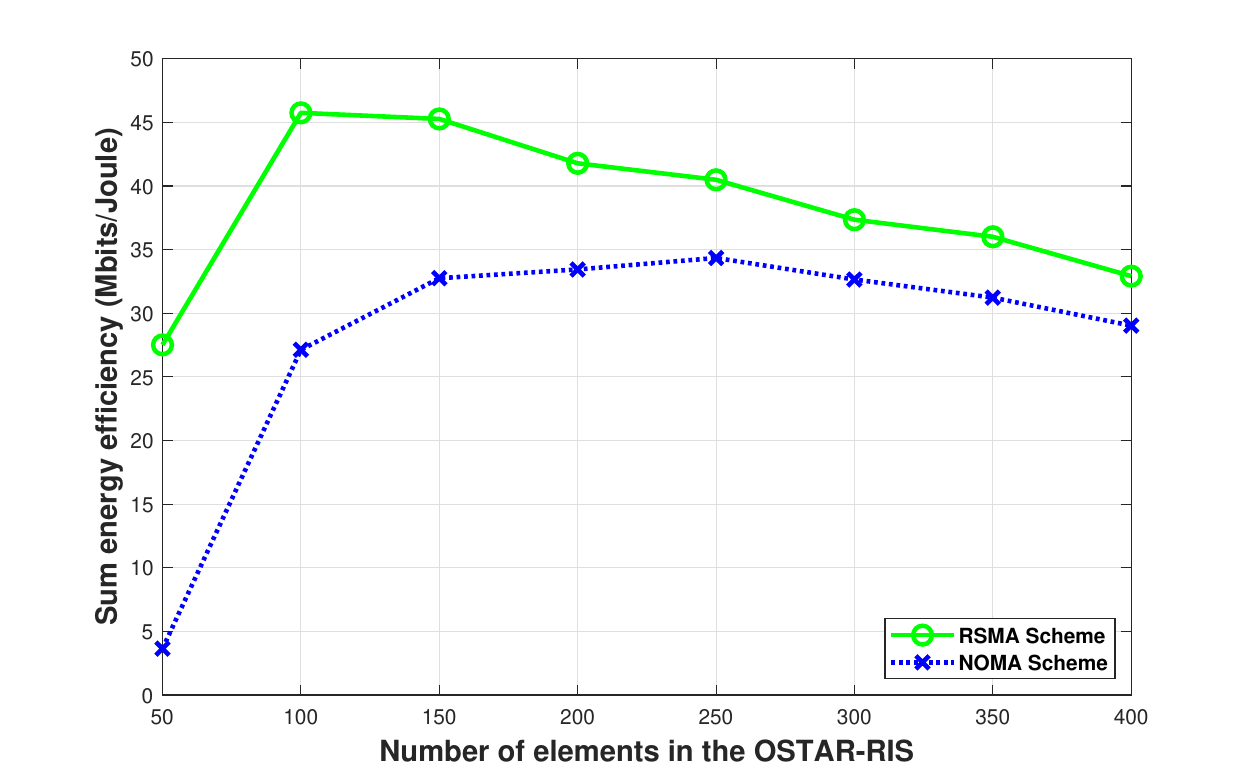}
        \label{fig: D_mirrors_vs_EE}
    }
    \caption{Proposed \gls{OSTAR-RIS}-assisted multi-user indoor \gls{VLC} system's performance versus number of \gls{OSTAR-RIS} elements. At $\lambda = 510$ nm and $p = 3$ Watts. }
\label{fig: D_mirrors_vs_R_P_EE}
\vspace{-2em}
\end{figure}

Fig.~\ref{fig: F_Variable_users} shows the sum rate performance of the proposed \gls{OSTAR-RIS}-assisted multi-user indoor \gls{VLC} system for the \gls{RSMA} scheme and the power-domain \gls{NOMA} scheme versus the number of users in the system. Clearly, the sum rate performance of the \gls{RSMA}-aided system outperforms the \gls{NOMA}-aided system. This is because the multiuser gain is higher in the \gls{RSMA} scheme than in the power-domain \gls{NOMA} scheme. 

In Fig.~\ref{fig: E_Fixed_NOMAalike_Random}, the sum rate performance of the proposed \gls{OSTAR-RIS}-assisted multi-user indoor \gls{VLC} system for the \gls{RSMA} scheme under three power allocation strategies is illustrated. Specifically, (i) equal power allocation strategy (adopted strategy) (i.e., distributing the power of the private streams equally), (ii) the \gls{NOMA}-alike power allocation strategy (i.e., distributing the power of the private streams using the same strategy utilized in the power-domain \gls{NOMA} scheme based on~\eqref{eq: allocated power ratio}) and (ii) the random power allocation (i.e., distributing the power of the private streams based on uniform random distribution). From this figure, it is evident that the \gls{RSMA} scheme achieves the best sum rate performance using (i). This is because the power allocation strategy in (ii) is designed for the \gls{NOMA} scheme and the power allocation strategy in (iii) might not utilize the whole portion of power dedicated to the private streams.

Fig.~\ref{fig: D_mirrors_vs_R_P_EE} demonstrates the sum rate, total consumed-power, and sum energy efficiency performance of the proposed \gls{OSTAR-RIS}-assisted multi-user indoor \gls{VLC} system for the \gls{RSMA} scheme and the power-domain \gls{NOMA} scheme versus the number of elements in the \gls{OSTAR-RIS}. Fig.~\ref{fig: D_mirrors_vs_R} reveals that as the number of elements in the \gls{OSTAR-RIS} increases the sum rate performance of both the \gls{RSMA} and the power-domain \gls{NOMA} schemes increases. It is worth mentioning that the percentage of this increase here is slower than the one in, for example, Fig.~\ref{fig: A_Tx_Power_vs_Rate}. This can be justified by the fact that introducing additional elements to a sizable \gls{OSTAR-RIS} results in minimal improvement. Fig.~\ref{fig: D_mirrors_vs_P} shows that the total consumed-power is identical for both the \gls{RSMA} scheme and the power-domain \gls{NOMA} scheme as the parameters related to both mentioned schemes are not factors in~\eqref{eq: total power consumption}. By examining Fig.~\ref{fig: D_mirrors_vs_EE}, it is apparent that the enhancement in the \gls{SEE} initially increases with the growing number of elements in the \gls{OSTAR-RIS}, but eventually declines. This pattern can be explained as follows: the system's sum rate (as shown in Fig.~\ref{fig: D_mirrors_vs_R}) logarithmically increases and the total consumed power of the proposed system (as depicted in Fig.~\ref{fig: D_mirrors_vs_P}) linearly increases while increasing the number of \gls{OSTAR-RIS} elements.

\section{Conclusion and Future Research Directions}
\label{Sec: Conclusion}
This paper proposes and investigates a novel multi-user indoor \gls{VLC} system that is assisted by \gls{OSTAR-RIS}. The proposed system combats challenging \gls{LoS} blockage scenarios resulting from (i) other users (i.e., blockers), (ii) walls within the indoor environment, and (iii) the orientation of the recipient's device. For the proposed system, both the sum rate and the \gls{SEE} optimization problems, for both the power-domain \gls{NOMA} and the \gls{RSMA} schemes, are formulated, solved, and evaluated. These optimization problems jointly design the roll and yaw angles of the mirror array-based \gls{OSTAR-RIS} elements as well as the refractive index of the \gls{LC}-based \gls{OSTAR-RIS} elements. The sine-cosine meta-heuristic algorithm is utilized to get the global optimal solution for both formulated multi-variate non-convex optimization problems. Detailed simulation results are provided to illustrate the superiority of the \gls{RSMA} scheme over the power-domain \gls{NOMA} scheme for the proposed system in terms of both the sum rate and the \gls{SEE} metrics while considering different network parameters such as the \gls{AP} optical transmit power, the wavelength of the transmitted light, the number of elements in the \gls{OSTAR-RIS}, and the number of served users. In addition, the effect of adopting different power allocation strategies on the sum rate performance of the \gls{RSMA} scheme is illustrated. 

This paragraph outlines some interesting future research directions that involve, (i) investigating the proposed \gls{OSTAR-RIS} in indoor-to-outdoor or outdoor-to-indoor \gls{VLC} scenarios and outdoor Vehicle-to-everything (V2X) communication scenarios, and (ii) validating the proposed OSTAR-RIS design using a test-bed experimental setup while considering several practical constraints, such as the illumination constraints, the interference from other light sources, the glare effect, the dimming support, the impact of RIS reflections on humans, the user mobility, the changes in indoor environments, and the dynamic network conditions.

\bibliographystyle{IEEEtran}
\bibliography{main}

\vskip -1\baselineskip plus 0fil

\begin{IEEEbiography}[{\includegraphics[width=1in,height=1.25in, clip,keepaspectratio]{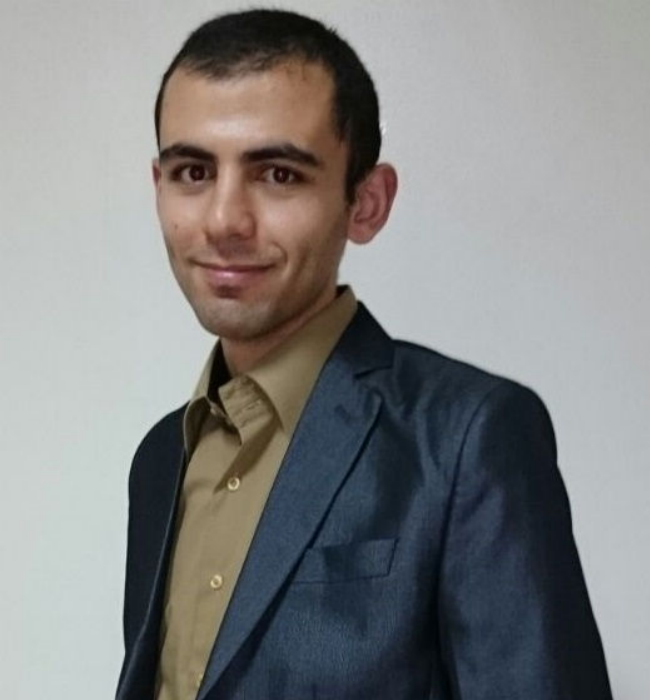}}]{\textbf{Omar Maraqa}} \hspace{0.01em} received the B.Eng. degree in Electrical Engineering from Palestine Polytechnic University, Palestine, in 2011, the M.Sc. degree in Computer Engineering and the Ph.D. degree in Electrical Engineering from King Fahd University of Petroleum \& Minerals (KFUPM), Dhahran, Saudi Arabia, in 2017 and 2022, respectively. He is currently a Postdoctoral Research Fellow with the Department of Electrical and Computer Engineering, at McMaster University, Canada. His research interests include optimization and performance analysis of emerging wireless communications systems.

Dr. Maraqa serves as a Technical Program Committee (TPC) Member for IEEE Vehicular Technology Conference, and a reviewer for several IEEE journals. He was recognized as an Exemplary Reviewer by IEEE Communications Letters in 2023.

\end{IEEEbiography}

\vskip -1\baselineskip plus 0fil

\begin{IEEEbiography}[{\includegraphics[width=1in,height=1.25in, clip,keepaspectratio]{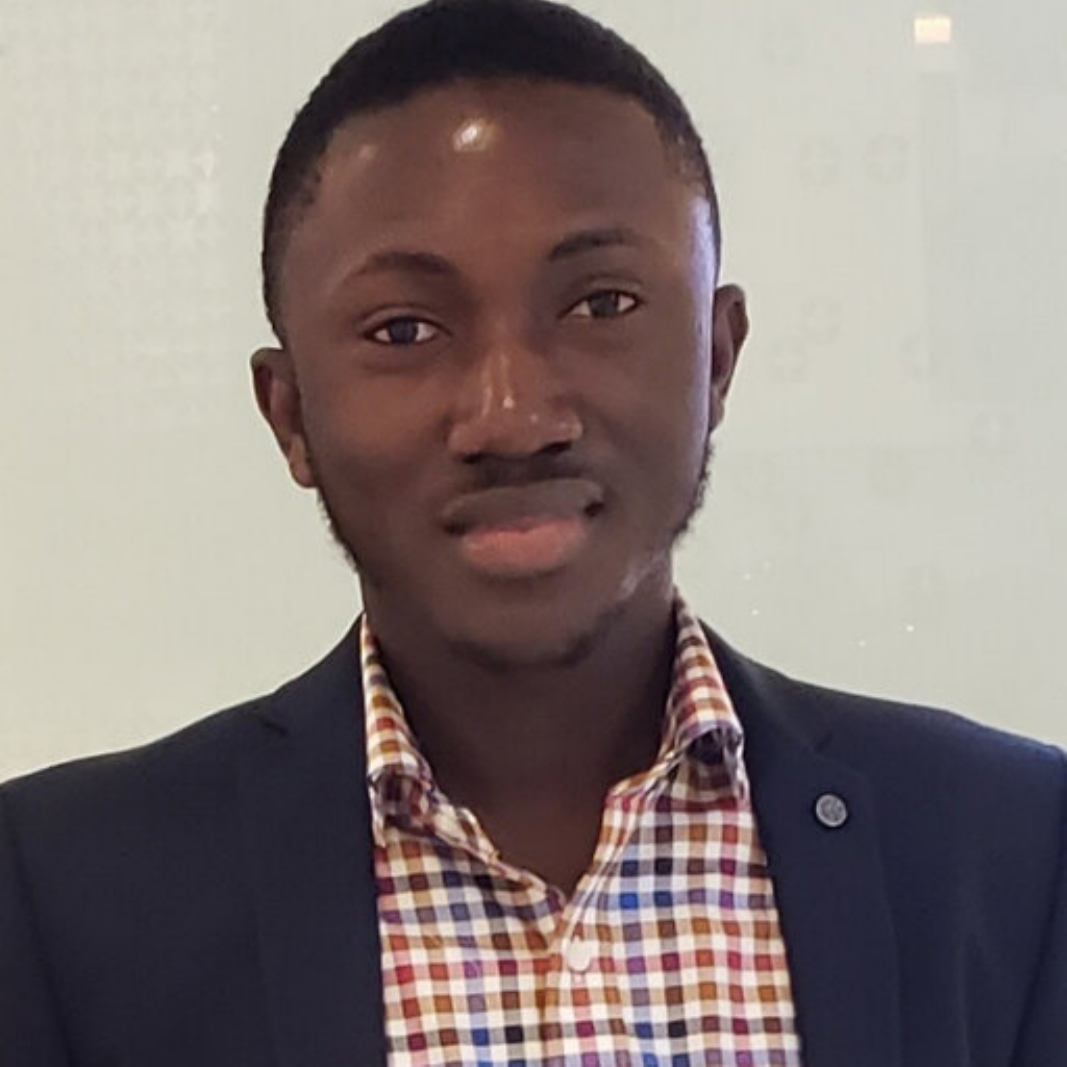}}]{\textbf{Sylvester Aboagye}} \hspace{0.01em} (Member, IEEE) received the B.Sc. degree (Hons.) in telecommunication engineering from the Kwame Nkrumah University of Science and Technology, Kumasi, Ghana, in 2015, and the M.Eng. and Ph.D. degrees in electrical engineering from Memorial University, St. John’s, NL, Canada, in 2018 and 2022, respectively. He received many prestigious awards including the Governor General’s Gold medal in graduate studies. He was a Postdoctoral Research Fellow with the Department of Electrical Engineering and Computer Science at York University, Canada, from January to December, 2023, and is currently an Assistant Professor with the School of Engineering,  University of Guelph, Canada. His current research interests include the design and optimization of multiband wireless communication systems, terrestrial and non-terrestrial integrated sensing and communication networks, and 6G and beyond enabling technologies. 

Dr. Aboagye serves as a Technical Program Committee (TPC) Member for IEEE Vehicular Technology Conference, and a reviewer for several IEEE  journals.  He was recognized as an Exemplary Reviewer by IEEE Communications Letters in 2023.
\end{IEEEbiography}

\vskip -1\baselineskip plus 0fil

\begin{IEEEbiography}[{\includegraphics[width=1in,height=1.25in, clip,keepaspectratio]{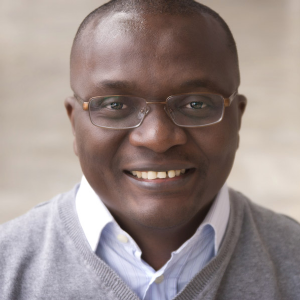}}]{\textbf{Telex M. N. Ngatched}} \hspace{0.01em} (Senior Member, IEEE) received the B.Sc. degree and the M.Sc. degree in electronics from the University of Yaoundé, Cameroon, in 1992 and 1993, respectively, the MscEng (Cum Laude) in electronic engineering from the University of Natal, Durban, South Africa, in 2002, and the Ph.D. in electronic engineering from the University of KwaZulu-Natal, Durban, South Africa, in 2006. From July 2006 to December 2007, he was with the University of KwaZulu-Natal as a Postdoctoral Fellow, from 2008 to 2012 with the Department of Electrical and Computer Engineering, University of Manitoba, Canada, as a Research Associate, and from 2012 to 2022 with Memorial University. He joined McMaster University in January 2023, where he is currently an Associate Professor. His research interests include 5G and 6G enabling technologies, optical wireless communications, hybrid optical wireless and radio frequency communications, artificial intelligence and machine learning for communications, and underwater communications. 

Dr. Ngatched serves as an Area Editor for the IEEE Open Journal of the Communications Society, an Associate Technical Editor for the IEEE Communications Magazine, and an Editor of the IEEE Communications Society On-Line Content. He was a recipient of the Best Paper Award at the IEEE Wireless Communications and Networking Conference (WCNC) in 2019. He is a Professional Engineer (P. Eng.) registered with the Professional Engineers Ontario, Toronto, ON, Canada.
\end{IEEEbiography}

\end{document}